\documentclass[modern]{aastex63}

\received{}
\revised{}
\accepted{September 1, 2021}

\submitjournal{ApJ}

\shorttitle{Signatures of Type III Solar Radio Bursts from Nanoflares}
\shortauthors{Chhabra et al.}

\graphicspath{{./}{figures/}}

\begin{document}

\title{Signatures of Type III Solar Radio Bursts from Nanoflares: Modeling}

\correspondingauthor{Sherry Chhabra}
\email{sc787@njit.edu}

\author[0000-0001-7754-0804]{Sherry Chhabra}
\affiliation{Center for Solar-Terrestrial Research, New Jersey Institute of Technology, Newark, NJ 07102, USA}
\affiliation{NASA Goddard Space Flight Center, Greenbelt, MD 20771, USA}

\author[0000-0003-2255-0305]{James A. Klimchuk}
\affiliation{NASA Goddard Space Flight Center, Greenbelt, MD 20771, USA}

\author[0000-0003-2520-8396]{Dale E. Gary}
\affiliation{Center for Solar-Terrestrial Research, New Jersey Institute of Technology, Newark, NJ 07102, USA}

\begin{abstract}

There is a wide consensus that the ubiquitous presence of magnetic reconnection events and the associated impulsive heating (nanoflares) is a strong candidate for solving the solar coronal heating problem.
Whether nanoflares accelerate particles to high energies like full‐sized flares is unknown. We investigate this question by studying the type III radio bursts that the nanoflares may produce on closed loops. 
The characteristic frequency-drifts that type III bursts exhibit can be detected using a novel application of the time-lag technique developed by \cite{2012ApJ...753...35V} even when there are multiple overlapping events. 
We present a simple numerical model that simulates the expected radio emission from nanoflares in an active region (AR), which we use to test and calibrate the technique. 
We find that in the case of closed loops the frequency spectrum of type III bursts is expected to be extremely steep such that significant emission is produced at a given frequency only for a rather narrow range of loop lengths. We also find that the signature of bursts in the time lag signal diminishes as: (1) the variety of participating loops within that range increases; (2) the occurrence rate of bursts increases; (3) the duration of bursts increases; and (4) the brightness of bursts decreases relative to noise. In addition, our model suggests a possible origin of type I bursts as a natural consequence of type III emission in a closed-loop geometry.

\end{abstract}


\section{Introduction} \label{intro}

Understanding the physics of the heating of the solar corona to several million K, three orders of magnitude higher than the observed surface temperature, is a challenge that has stirred solar research for decades. The high temperature of the active region corona is maintained by an input energy flux rate of $\approx10^7$ ergs cm$^{-2}$ s$^{-1}$, and that of the quiet Sun by $\approx3\times10^5$ ergs cm$^{-2}$ s$^{-1}$, which is the energy flux required to offset the radiation losses, mainly in the form of extreme ultraviolet (EUV) and X-rays \citep{1977ARA&A..15..363W}. One of the plausible mechanisms that may be responsible for this comprehensive phenomenon is the energy produced from dissipation of waves (Alfvén or magneto-acoustic). This is referred to as alternating current (AC) heating \citep[see][]{1947MNRAS.107..211A, 1958ApJ...128..664P,1961ApJ...134..347O}. On the other hand, \cite{1983ApJ...264..635P,1988ApJ...330..474P} argued that small scale photospheric convective motions stress the magnetic strands in the corona, which then break via magnetic reconnection to give impulsive energy releases known as ‘nanoflares.’ This is known as direct current (DC) heating. Further investigations have revealed that nanoflares are ubiquitous in the solar atmosphere, and despite releasing relatively low energy as compared to a full-sized flare, may still heat the corona collectively due to their sheer numbers \citep[][and references therein]{1988ApJ...330..474P, 2006SoPh..234...41K, 2015RSPTA.37340256K}. A major challenge we face is the lack of direct detection of individual nanoflares, due in part to their small amplitude and in part to the confusion from line-of-sight overlap of the optically-thin structures. Therefore, the existence and properties of nanoflares must be inferred from their collective effects. A variety of different methods have been used \citep{hinode2019JAK}.

During a nanoflare life-cycle, the EUV emission from the heating phase is much less than that from the slower cooling phase \citep{2011ApJS..194...26B}. This property of nanoflares leaves a unique signature in multi-wavelength observations. As the plasma cools, the loop appears first in a hot channel, subsequently showing up in cooler channels which, in turn, reach their peak intensities with some time delay. The powerful technique developed by \cite{2012ApJ...753...35V} can identify cooling patterns in large ensembles of loops by detecting even the minutest variability in light curves. They use high-cadence observations from the Atmospheric Imaging Assembly onboard the Solar Dynamic Observatory \citep[SDO/AIA;][]{2012SoPh..275...41B, 2012SoPh..275...17L} spacecraft to measure the time-lag between two coronal channels. This is accomplished by computing the cross correlation of the light curves with different amounts of imposed temporal offset and determining which offset maximizes the correlation \citep[see][]{2012ApJ...753...35V,2017ApJ...842..108V}. The technique not only detects time lags that can be identified by eye in observationally distinct loops, but also works well when there are countless overlapping coronal strands in observations of the diffuse corona, where the time lags are not obvious to the eye. \cite{2013ApJ...771..115V} simulated the composite emission from 10,000 strands heated randomly by nanoflares, as expected in a real solar observation. The light curves exhibit only small fluctuations, yet the time-lag technique correctly identifies the cooling that is known to be present in the simulations. 

Full-sized flares are extremely efficient at accelerating particles to high energy. Whether this is also true of nanoflares is unknown \citep{Vievering2020}. A leading theory of particle acceleration involves collapsing plasmoids \citep{2006Drake}. The theory predicts that the efficiency of acceleration depends on the magnetic field geometry - in particular, whether there is a strong guide field component \citep{2015PhPl...22j0704D}. Because nanoflares and flares have different geometries, determining whether particle acceleration occurs in nanoflares would be an important test of the theory.  Instinctively, we turn to hard X-ray observations to answer this question. Such observations place a rather low upper limit on the quantity of highly nonthermal electrons. However, large numbers of mildly
nonthermal electrons are not ruled out in active regions, because their emission would
be dominated by much brighter thermal emission in HXR, rendering the nonthermal
component undetectable. For quiet Sun, although the temperatures are low enough
that no thermal emission is observed to dominate HXR energies, nonetheless, the
lack of sensitivity of the current instruments does not place any limits on the mildly
energetic particles. \citep{Hannah2007, Hannah2010, Ishikawa2019}. Therefore, understanding particle acceleration from nanoflares requires a different approach. 

\begin{figure}[h!]\centering
\includegraphics[width=0.5\textwidth, clip]{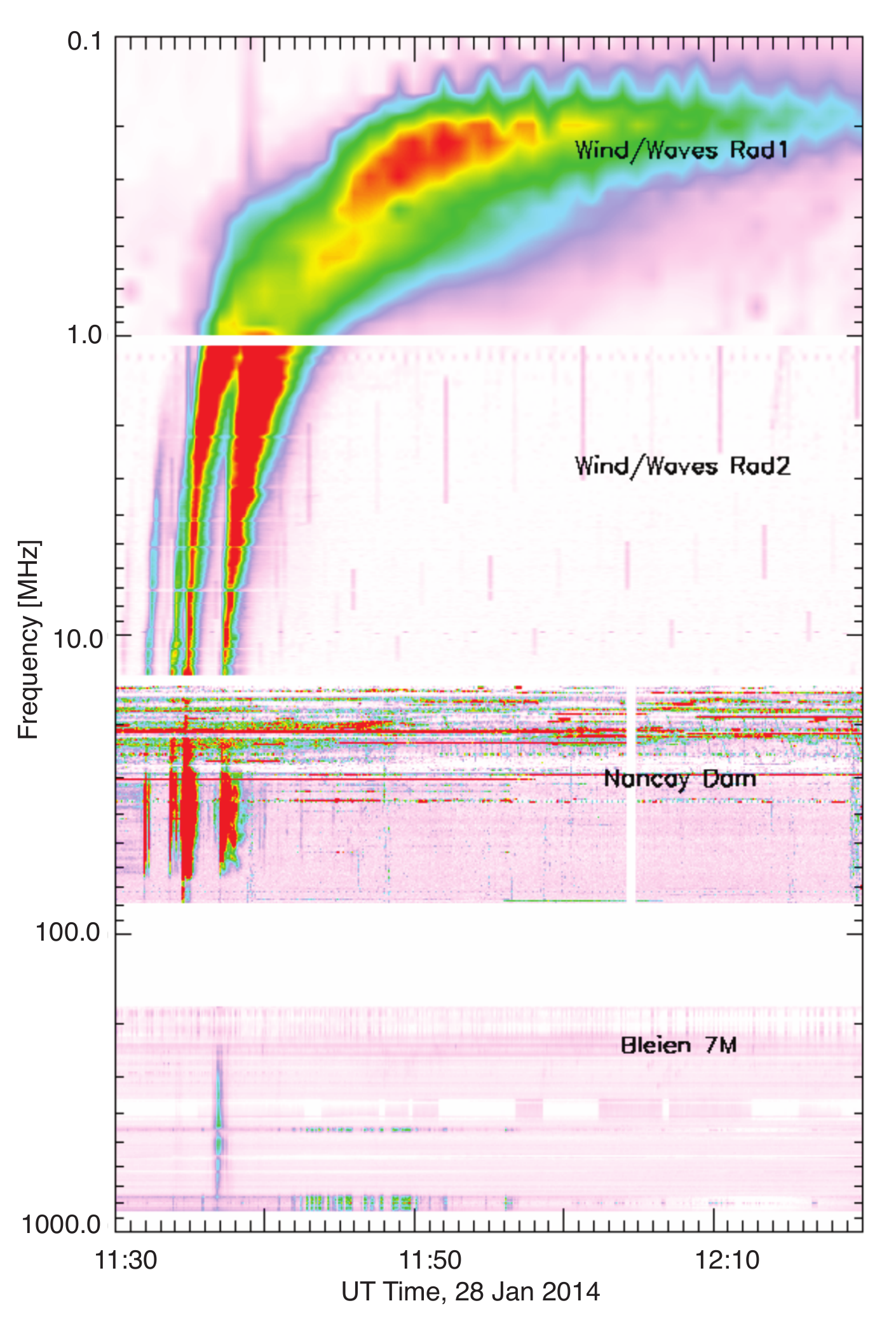}
\caption{\label{typeIII} Composite dynamic spectrum from the Bleien telescope, the Nancay Decametre Array, WIND/WAVES RAD2, and RAD1 exhibiting traditional interplanetary type III radio bursts observed on January 28th, 2014 (\cite{2014RAA....14..773R}, Figure 1). Reprinted with permission from Research in Astronomy and Astrophysics} 
\end{figure}

Radio emission on the Sun is not constrained by these limitations. Type III radio bursts (coronal and interplanetary) especially are an important tool for understanding accelerated electron beams. These bursts, often associated with flares, have been observed to be prevalent in a broad range of frequencies (10 kHz--1 GHz). Energetic electron beams of semi-relativistic speeds ($\approx\,$0.1--0.5\,$c$) produce a bump-on-tail instability as they propagate along the magnetic field, creating electrostatic Langmuir waves at the local plasma frequency, \(\textup{$\nu_p$} = 8980\,\sqrt[]{n_e}\) Hz \citep{1958SvA.....2..653G}. Non-linear particle-wave and wave-wave interactions generate electromagnetic emission at this local plasma frequency and its second harmonic, respectively. Because there is typically a density gradient along the beam's path, they show a characteristic frequency drift in the radio dynamic spectrum \citep[][and references therein]{1950AuSRA...3..387W,2014RAA....14..773R}.

Figure \ref{typeIII} shows an example of traditional type III radio bursts as the particles move away from the Sun into interplanetary space. The dynamic spectrum combines data from the Bleien telescope covering the frequency range of 900--200~MHz \citep{2009SoPh..260..375B}, the Nancay Decametre Array \citep{2000GMS...119..321L} covering the range of 80--15~MHz, and the RAD1 and RAD2 receivers onboard the WIND spacecraft \citep{1995SSRv...71..231B} covering 14--0.1~MHz.  The frequency axis is inverted (decreasing upward) to give the impression of the electron beam propagating away from the Sun. As the electrons propagate outward along open field lines, they encounter decreasing density, and the emission is first detected at a higher frequency, followed by progressively lower frequencies. The time delay between the two defines the frequency drift rate of the burst. Sometimes beams travel up and then down along closed field lines (i.e., loops) where they encounter first decreasing then increasing density, which is manifested as inverted ‘U’ or ‘J’ shaped structures in the dynamic spectrum, with frequency drifts in both directions.

While these traditional type IIIs are observed as individual bursts, in groups or sometimes as type III storms with hundreds of bursts occurring over a period of a few hours, the ever-present nanoflares may produce hundreds or thousands of bursts per second, which will not be identifiable as individual burst features in the dynamic spectrum. Rather, they may present themselves as a quasi-continuous ‘radio haze.’ The collective emission from these small bursts will not be as intense, and small but real fluctuations may be misinterpreted as mere noise. One such example is seen in the recent work by \cite{Mondal2020} that observed ubiquitous weak radio bursts from the quiet solar corona. Moreover, the emission from these bursts coming from the closed corona will exhibit frequency drifts in both directions, similar to those seen in the stronger ‘U’ or ‘J’ bursts noted above.

In this paper, we attempt to identify signatures of type III radio bursts that may be produced by nanoflares using the time-lag technique. Similar to correlating light curves in pairs of EUV channels, we will correlate light curves in pairs of radio frequencies. The radio drift of overlapping type III bursts shows up like the EUV cooling signature from overlapping magnetic strands, although on much shorter timescales. 
We construct a simple numerical model simulating emission produced by these type III bursts at different frequencies, which are then cross-correlated to measure time-lags for each frequency pair. We add increasing complexity to the model to more closely approximate the real corona, and at each step we evaluate the efficacy of the time-lag technique. Section 2 gives the detailed formalism for the loop model along with the methodology used to populate loops with nanoflares and simulate type III emission. We discuss our results in section 3. The Appendix summarizes various additional factors  that may affect the cross-correlation results. Subsequent papers are planned that apply these results to observations at these frequencies from the FIELDS experiment on Parker Solar Probe, the Very Large Array (VLA), and the Low-Frequency Array (LOFAR).  

\section{Modeling} \label{modeling}

\subsection{Loop Model Formulation} 
The basic building block of our model is a symmetric loop in static equilibrium. Following \cite{2010ApJ...714.1290M}, we obtain solutions to the one-dimensional energy equation:

\begin{equation}\label{eq1}
\frac{d}{ds}\left(\kappa_o T^{5/2}\frac{dT}{ds} \right) + Q - P^2_o \chi_o T^{-(2+\gamma)}  =  0
\end{equation}
where $s$ and $T$ are the spatial coordinate and temperature along the loop, $\kappa_o(=1.1\times10^{-6}$ erg cm$^{-1}$ s$^{-1}$ K$^{-7/2}$) is the coefficient of thermal conductivity, $Q$ is the volumetric heating rate, $P_o$ is the gas pressure, taken to be constant along the loop due to the large gravitational scale height, and $\chi_o$(=$10^{12.41}$) \& $\gamma(=0.5$) \citep{Martens2000} are parameters of the optically-thin radiative loss function in c.g.s. units. The heating function is taken to have a power-law dependence on pressure and temperature:

\begin{equation}\label{eq2}
    Q = H P^\beta_oT^\alpha
\end{equation}
which is further reduced to 

\begin{equation}\label{eq3}
    Q = H 
\end{equation}
assuming $\alpha$ = $\beta$ = $0$ for a uniformly heated loop, where $H$ is a constant of proportionality. 

Although nanoflare heating is inherently impulsive, we believe it is reasonable to assume steady heating and static equilibrium for this initial evaluation of the feasibility of detecting overlapping type III bursts. Furthermore, nanoflares are believed to recur on individual field lines with a range of repetition frequencies \citep[][and references therein]{hinode2019JAK}. High-frequency nanoflares are effectively similar to steady heating, and loops heated by low-frequency nanoflares spend much of their time in a phase that is not greatly different from static equilibrium conditions.

Static equilibrium loops obey well-known scaling laws. With uniform heating, we have  \cite[][equations~28 \& 30]{2010ApJ...714.1290M}:

\begin{equation}\label{eq4}
  P_oL = T_a^3\left(\frac{\kappa_o}{\chi_o}\right)^{1/2} \frac{\sqrt{2}}{5} B\left(\frac{6}{5},\frac{1}{2}\right)  
\end{equation}
 and 

\begin{equation}\label{eq5}
   Q \approx \frac{1}{2}\kappa_o \frac{T_a^{7/2}}{L^2}
\end{equation}

\noindent where $L$ is the loop half-length, $T_a$ is the temperature at loop apex, and $B$ is the beta function. 

Magnetic strands become entangled and braided as they are churned by chaotic photospheric motions. One promising idea is that nanoflares occur when the angle between adjacent misaligned strands reaches a critical value. As discussed in \citet{2000ApJ...530..999M}, this leads to a volumetric heating rate that scales with the loop length according to $Q~=~cL^{-3}$, where $c$ is a constant. Adopting this dependence in the scaling laws (Equation \ref{eq5}) above, we obtain a value for constant, $c$, for a typical loop length of $30,000$ km with an apex temperature of $2$~MK. Further, using the IDL routine BETALOOP.PRO developed by \citet{2010ApJ...714.1290M}, we can construct density and temperature profiles for loops with a variety of different lengths. Note that the temperature and density profiles thus obtained come from the above model and satisfy the energy balance equation (Equation \ref{eq1}). Example density profiles for three loops of different lengths (electron density, $n$, vs. distance, $s$, along the loop) is shown in Figure \ref{int_cartoon}a. As is characteristic of all loops, the  gradients are shallow in the corona and steep in the transition region (TR) near the base. The temperature profile has these same properties, being essentially the inverse of the density profile due to the constant pressure.

\subsection{Type III logistics} \label{logistics} 

As already noted, a beam of energetic electrons will generate radio emission at the local plasma frequency, $\nu_p \propto n^{1/2}$. This involves complex nonlinear interactions among the ambient ions or ion-acoustic waves and Langmuir waves that are generated by the beam. Detailed simulations of these processes have been performed in the context of the magnetically open corona and solar wind \citep{Kontar2009, Ratcliffe2014, Reid2018}.
We do not attempt that here. Rather, we assume that the type III emission from a given location turns on, maintains a constant brightness, and turns off during the time that the beam is passing that spot. Although, substantial work has been done to understand the effects of density fluctuations, scattering and refraction on the brightness temperature and source sizes of traditional type IIIs \citep[][]{steinberg1984,Thejappa2008, kontar2019}, how the brightness may depend on parameters such as the density and energy of the beam, the ambient density, or the strength of the magnetic field are largely unknown . We therefore assume that the emissivity (emission rate per unit volume) is the same for all events and all loop positions. This approach is consistent with our goal of demonstrating the feasibility of the technique in this initial study.  

\begin{figure}[h!]\centering
\includegraphics[width=1.0\textwidth, clip]{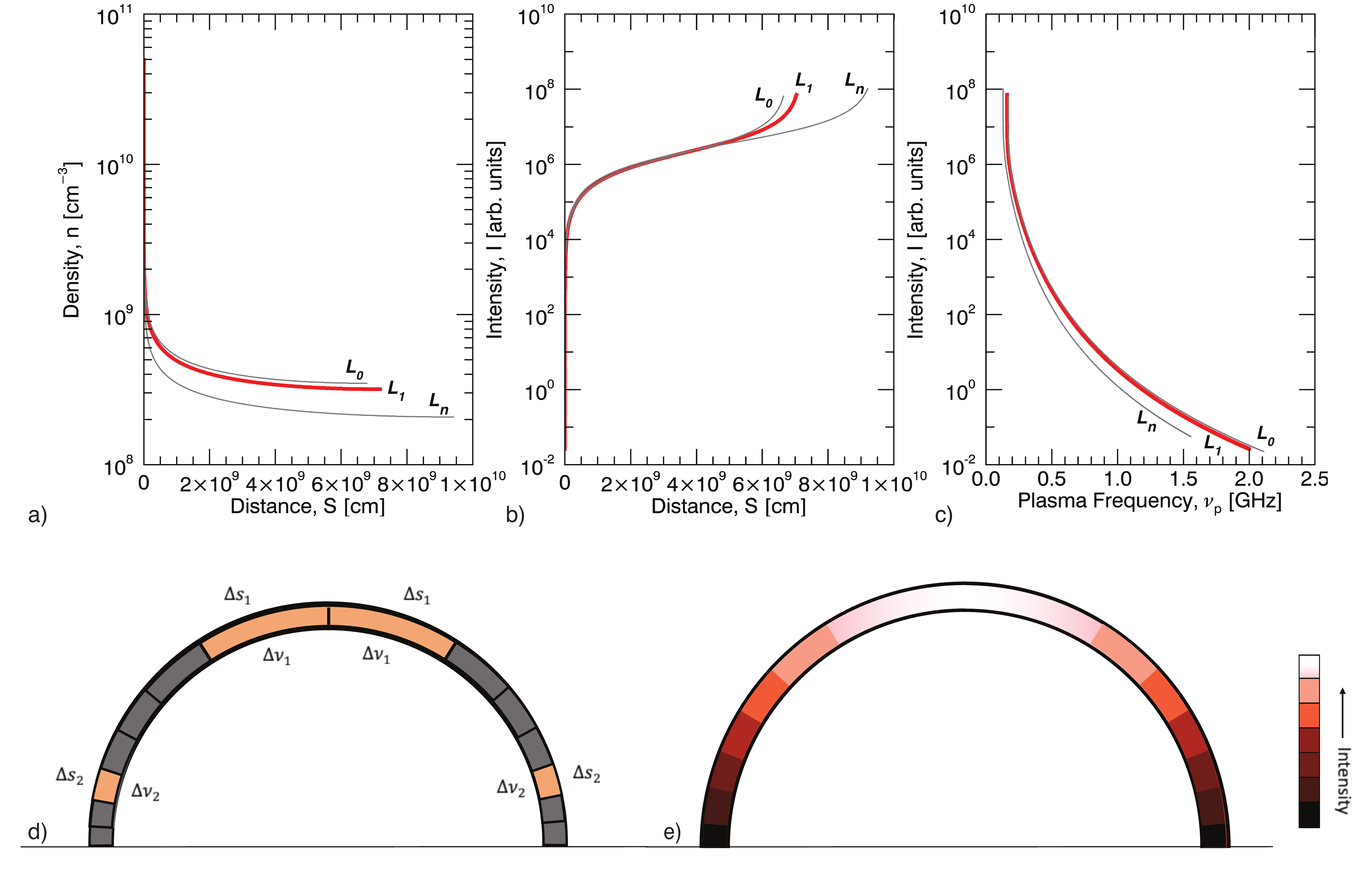}
\caption{\label{int_cartoon}Loop properties for three different loop lengths: a) Density profile for three loops: $L_0, L_1$ and $L_n$ as marked in Fig. \ref{IntnCCOP}, as a function of distance, b) Relative Intensity, $I$, in frequency bins of equal size as a function of distance along the loop for a type III burst, c) Relative Intensity as function of local plasma frequency along the loop. d) $\Delta\nu_i$ bins of same width occupy a much larger volume of the loop in the upper corona compared to the TR. e) Intensity variation for a burst vs. location in the loop. Note that this is not a rigid scale intended to show intensity along the loop in the figures that follow. It is only meant to illustrate the trend of intensity variation along the length of any loop.}
\end{figure}

Radio spectral observations usually sample the emission in frequency bins, or channels, of equal size, $\Delta\nu$. We do the same when generating synthetic spectra from our models. Because density varies along the loop, each frequency bin corresponds to a small range of densities, $\Delta n$, which in turn corresponds to a finite section of the loop, $\Delta s$, as shown schematically in Figure~\ref{int_cartoon}d. We refer to this as a volume element, under the assumption of constant cross-sectional area \citep{2000Klimchuk}.

Depending on the beam duration and hence the beam length, volume elements may be partially or fully filled by the beam. Longer elements near the apex tend to be partially filled, while shorter elements in the lower legs tend to be fully filled. We assign an intensity to frequency $\nu_i$ that is proportional to either the beam length $L_b$ or the length of the corresponding volume element $\Delta s_i$ for elements that are longer and shorter than the beam, respectively. Elements are centered on $s_i$ where the local plasma frequency is $\nu_i$. We also take into account the finite time required for the front and back of the beam to traverse the element; the light curve at a given frequency ramps up linearly as the beam enters the element, has a flat section, and ramps down linearly as the beam exits. The frequency bin size $\Delta \nu_i$ is the same for all frequency bins (all $i$). 

Note that for a sufficiently long beam length $L_b$, such that all volume elements $\Delta s_i$ along the loop are shorter than $L_b$, the intensity at all frequencies $\nu_i$ will be directly proportional to their respective volume elements $\Delta s_i$.

The size of a volume element and its associated intensity vary inversely with the local density gradient, as shown in Appendix \ref{intensity}. Gradients are very small near the loop apex and increase steadily down the legs, becoming very large near the footpoints. Consequently, the intensity is a strong function of frequency (Figure~\ref{int_cartoon}c). Lower frequencies come from high in the loop and are bright, while higher frequencies come from low in the loop and are faint. This is indicated in Figure~\ref{int_cartoon}b,c. Appendix \ref{intensity} shows that for a constant conductive flux, which is a crude approximation to an equilibrium coronal loop, the intensity varies as $n^{-4}$ and therefore as $\nu^{-8}$. The very steep spectrum has important implications, as we discuss below. Keep in mind that this is the spectrum of a single loop. The composite spectrum from many loops with different density profiles is much more uniform. 

Referring to Figure \ref{int_cartoon}, consider two frequencies $\nu_1$ and $\nu_2$ occurring high and low in the loop, respectively. The frequency bins  $\Delta \nu_1$ and $\Delta \nu_2$ centered on $\nu_1$ and $\nu_2$ have the same width, but the corresponding volume elements $\Delta s_1$ and $\Delta s_2$ are different; $\Delta s_1 > \Delta s_2$ because of the smaller density gradient. The intensity is therefore brighter at $\nu_1$. The shading in Figure \ref{int_cartoon}e indicates the total emission coming from each volume element . The differences are due entirely to the variations in volume. Our model assumes that emissivity is uniform along the loop. Similarly, the intensity plotted in Figure \ref{int_cartoon}b is not the emission per unit volume at position $s$, but rather the intensity of the volume element that is centered at $s$. 
 
We now explore the light curves (intensity versus time) expected when a beam of energetic electrons propagates along the loop. Different frequencies will light up at different times. By cross correlating the light curves with different imposed temporal offsets, we generate a Cross-COrrelation Power Spectrum (CCOPS) \footnote{The term ``Power Spectrum'' refers to the variation of the amplitude of the cross-correlation coefficient as a function of time-offset and should not be confused with the Fourier transform of the auto-correlation function.}. We want to know whether the CCOPS reveals a signature of the beam that might be used to identify type III bursts from nanoflares. We start with very simple scenarios and add increasing complexity to more closely mimic the actual Sun.

\subsection{Single-Loop Model} \label{single loop}

\begin{figure}[h!]\centering
\includegraphics[width=1.0\textwidth, clip]{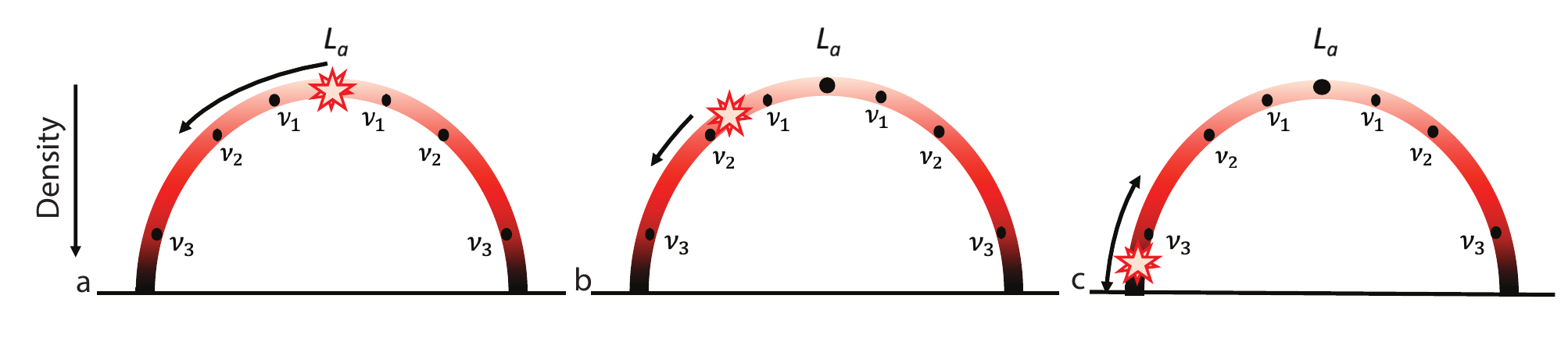}
\caption{\label{cases_cartoon}Case A: nanoflare at the loop top with particles moving downward to the foot-point; Case B: nanoflare high in the loop leg with particles moving downward only; Case C: nanoflare low in the loop leg with particles moving in both directions. }
\end{figure}

\begin{figure}[h!]\centering
\includegraphics[width=0.6\textwidth, clip]{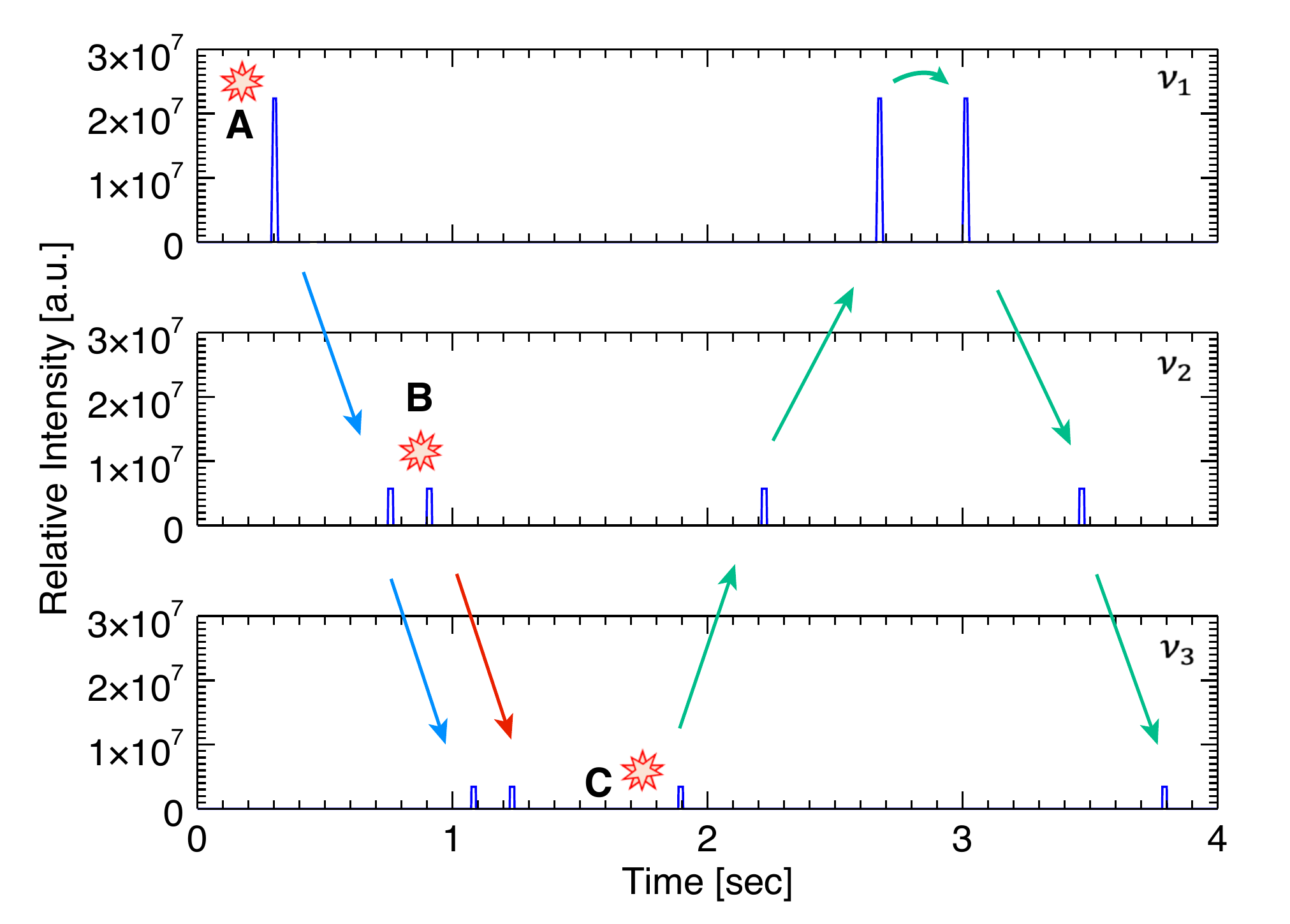}
\caption{\label{cases_lc}Light curves for the three chosen frequencies, showing time offsets as per Cases A, B, and C. The ‘red star’ marks the relative position  along the loop where each nanoflare occurred. }
\end{figure}

\textit{Model 1}: First consider a single loop with half-length $L$, as depicted in Figure \ref{cases_cartoon}. Sample frequencies $\nu_1, \nu_2,$ and $\nu_3$ correspond to the positions indicated. The loop is symmetric, so each frequency is present in both legs. We now imagine that a nanoflare occurs at three different locations, which we refer to as cases A, B and C. These produce the light curves shown in Figure \ref{cases_lc}, ordered so frequency increases from top to bottom to reflect the relative heights within the loop. The red 'stars' indicate the position and timing of the three different nanoflares.

\textit{Case A}: The nanoflare occurs at the loop apex ensuring all particles only propagate downward (Fig.~\ref{cases_cartoon}a). The ejected beam of electrons is assumed to be mildly non-thermal with a fixed energy of $2$~keV, hence moving with a constant velocity of $2.65\times 10^9$ cm/s ($\sim0.1c$). The duration of the beam represents the time that the emission stays 'on' at a given location, and is chosen to be $80~$ms for these examples. As the beam propagates along the field, we expect to see emission first at $\nu_1$ followed by $\nu_2$ and then $\nu_3$. The features in the light curves joined by the blue arrows in Figure~\ref{cases_lc} clearly demonstrate this systematic behavior. 

\textit{Case B}: Here we assume that the nanoflare occurs partway down the left leg and that particles propagate downward only. Following from the sketch for case B (Fig. \ref{cases_cartoon}b), it is evident that only $\nu_2$ and $\nu_3$  will show any emission (features joined by the red arrows) since the nanoflare lies below $\nu_1$ and the downward moving particles never pass that location. Emission appears in  $\nu_2$ and $\nu_3$ with the same delay as in Case A. We classify this as a positive delay, because the higher frequency occurs after the lower frequency.

\textit{Case C}: The nanoflare now occurs lower in the loop leg and particles propagate in both directions (Fig. \ref{cases_cartoon}c). Downward propagating particles produce no emission at any of the three frequencies. Upward propagating particles produce emission in $\nu_3$ followed by $\nu_2$ and then $\nu_1$ (green arrows in the figure) as they travel up the leg. These are negative delays because the higher frequency occurs first. As the particles pass the apex and travel down the opposite leg, they produce emission sequentially in $\nu_1$, $\nu_2$, and $\nu_3$, the reverse order. These are positive delays.  This reversing of the frequency drift is the same effect observed in U-type bursts mentioned earlier.

\begin{figure}[h!]\centering
\includegraphics[width=1.0\textwidth, clip]{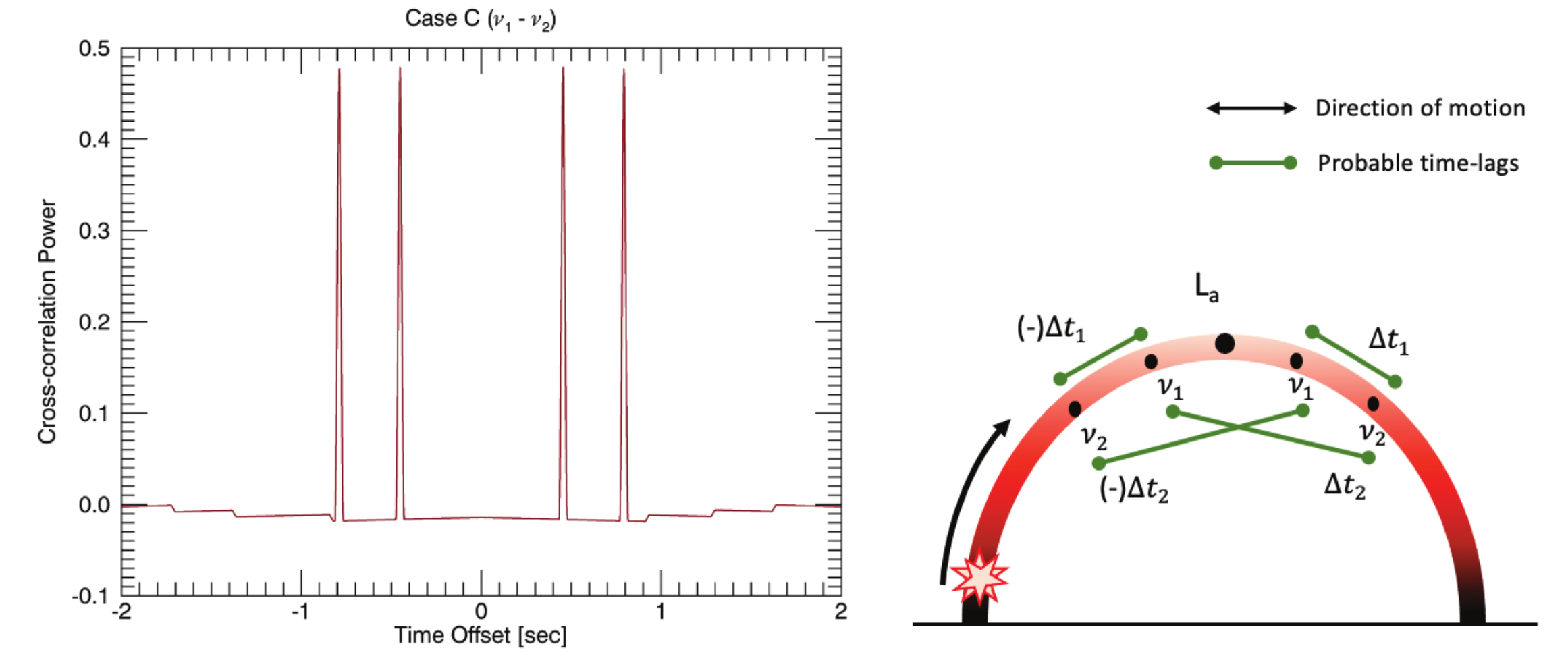}
\caption{\label{sample_CCOP}Left: CCOPS for $\nu_1 - \nu_2$ from light curves obtained using a generalization of case C. Right side demonstrates the multiple time-lags that peak in the CCOPS}
\end{figure}

The time-lag technique easily identifies the expected delays in these simulations with each nanoflare corresponding to cases A, B, and C. The CCOPS has spikes at the temporal offsets discussed above. However, in addition to finding correlations between two frequencies occurring in the same leg, the technique also finds correlations with longer delays when the burst arrives at the opposite leg. This is illustrated in the example CCOPS in Figure \ref{sample_CCOP} corresponding to case C from above.
We consider frequencies  $\nu_1$ and $\nu_2$ that occur at higher altitudes. As indicated in the sketch on  the right, there are four combinations of  $\nu_1$--$\nu_2$ positions.  $\nu_2$ on the left side of the loop pairs with both $\nu_1$ on the left and $\nu_1$ on the right. Both delays are negative, but the first is shorter than the second. $\nu_1$ on the left pairs with $\nu_2$ on the right, producing a long positive delay. Finally, $\nu_1$ on the right pairs with $\nu_2$ on the right, producing a short positive delay. It is clear that the same pattern will hold for nanoflares occurring near the right footpoint, with the beam traveling right to left. In general, we expect the CCOPS to have peaks at four temporal offsets:  positive and negative short delays and, positive and negative long delays. This is exactly what we find in our simulation, as shown on the left of Figure \ref{sample_CCOP}. There will always be peaks at $\pm \Delta t_1$ and $\pm \Delta t_2$ whenever some nanoflares occur at altitudes below the higher frequency, $\nu_2$.

The situation is modified if nanoflares occur only at higher altitudes in the loop. There will be only positive delays, both short and long, if the nanoflares are distributed at locations above the higher frequency, and only positive short delays if they are restricted to the very top of the loop, above the lower frequency.

The above examples concern a single loop. In reality, of course, many loops with different density profiles are observed at the same time, with nanoflares occurring randomly along the loop lengths, and this adds enormous complexity to the CCOPS. There are estimated to be $\sim$500,000 individual strands (unresolved loops) in a typical AR \citep{2015RSPTA.37340256K}. The characteristic delay between successive nanoflares in a given strand is $\sim$1000 s \citep{hinode2019JAK}, implying an occurrence rate of  $\sim$500~nanoflares/s across the AR. Even with high spatial resolution observations, the line of sight passes through many loops of differing lengths, heating rates, and density structures. We expect the CCOPS to contain a huge number of peaks, which merge together to produce a smooth spectrum. Meaningful patterns in the envelope may nonetheless persist, as we now discuss. We first consider the effect of multiple loops and then consider the effect of the type III burst occurrence rate.

\begin{figure}[h!]\centering
\includegraphics[width=0.95\textwidth, clip]{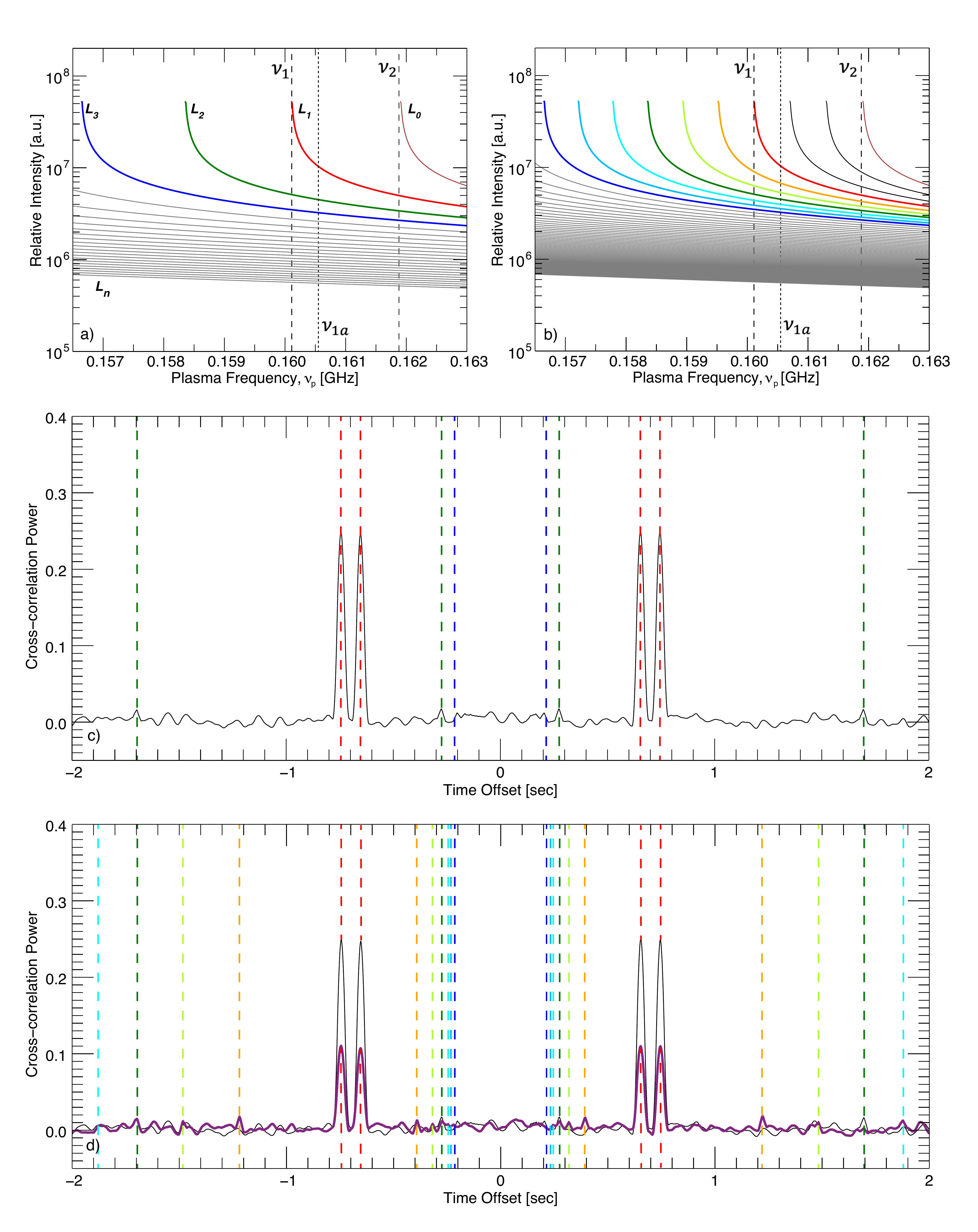}
\caption{\label{IntnCCOP} a. Low-density ($\eta = 1$) distribution of loops generated for the chosen loop-length range from frequencies $\nu_1$ and $\nu_2$. b. Loop distribution with 3 times the lowest loop density ($\eta = 3$). c. CCOPS for loop density $\eta = 1$ with color-coded dashed lines overlaid to represent the expected time-offsets for each loop. d. CCOPS for loop density $\eta = 3$ (purple line), with color coded dashed-lines representing expected time-offsets for the additional loops as well. The CCOPS for loop density $\eta = 1$ is overlaid illustrating the drop in power due to the addition of new loops.}
\end{figure}

\subsection{Multiple Loops} \label{multiple loops}

Loop half-lengths in an AR typically vary between $\sim$10,000--150,000~km \citep{2000ApJ...530..999M} and can have apex temperatures between $\sim$1--4~MK \citep{Warren2012}, giving a wide range of density profiles and therefore a broad band of plasma frequencies that can be observed. However, the precipitous fall in intensity as a function of frequency shown in Figure~\ref{int_cartoon}c has important implications. Systematic time delays between any two frequencies are only expected for the emissions coming from a single loop. Emissions from different loops are physically uncorrelated and therefore have random delays. The rapid drop in intensity for a single loop means that we expect an observable signal from that loop only for frequencies that are closely spaced. If the spacing is too great, at least one of the frequencies will be too weak to produce a meaningful signal. Because we require a reasonable signal in both frequencies, we need concern ourselves only with loops having a rather narrow range of apex densities (uniquely determined by their lengths in our simplified model). All other loops will be faint at these frequencies.
This also shows that in order to consider emission from all loops in an active region, multiple different pairs of frequencies will need to be considered.

\subsubsection{Loop Distribution} \label{loop distribution}

Consider the spectra of several loops with equally spaced lengths $\Delta L$ shown in Figure \ref{IntnCCOP}a. Note that the intensity scale is logarithmic. Four of the loops are color coded and labeled $L_0$ through $L_3$. The $\Delta L$ increment of loop lengths is determined by the following choice of frequencies. Frequency $\nu_1$ occurs just below the apex of $L_1$ (red),
and frequency $\nu_2$ occurs a short distance down the leg, where the intensity is reduced by a factor of 10. The plasma frequency at the apex of $L_0$ (brown) is slightly larger than $\nu_2$, so neither frequency occurs in that loop. The longest loop shown (labeled $L_n$ at the bottom) is approximately 100 times fainter than $L_1$ at frequency $\nu_1$. We assume that all loops longer than this are so faint that they can be safely ignored in our simulations.

 \begin{figure}[h!]\centering
\includegraphics[width=1.0\textwidth, clip]{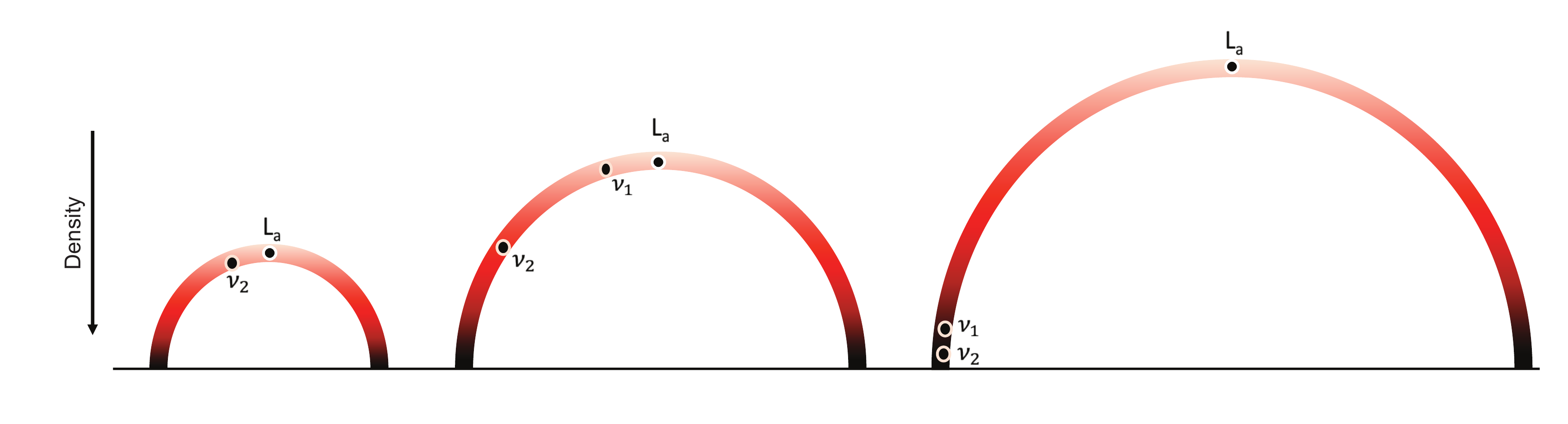}
\caption{\label{difflenloops}Relative position of frequencies along loops of varied lengths. As the loop become shorter from right to left, the density at the loop-top increases such that only one of the frequencies, $\nu_2$ exists along the shortest loop. For a loop even shorter, the density will increase further such that neither of the frequencies exist along the loop anymore.}
\end{figure}

\begin{figure}[h!]\centering
\includegraphics[width=0.6\textwidth, clip]{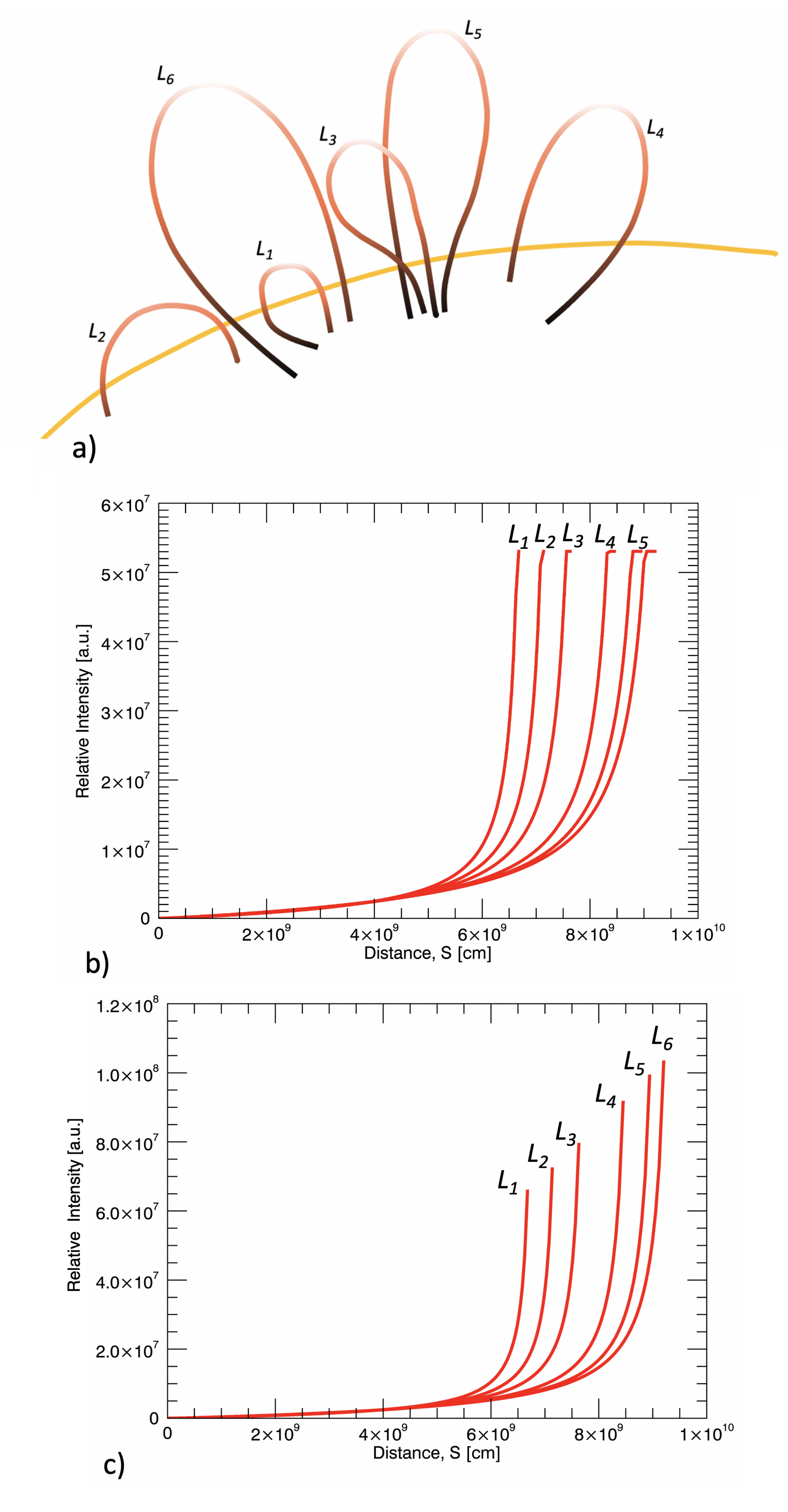}
\caption{\label{orientation} a) Cartoon showing multiple loops in an AR with different orientations and lengths. (b-c) Corresponding plots for Relative Intensity as a function of distance along the loop $I~vs~s$ for two cases, with short beam length b) and long beam length c).}
\end{figure}

Figure \ref{difflenloops} is a schematic representation of the loops. The middle loop is similar to $L_1$, with both frequencies being relatively bright. The long loop would correspond to one of the unlabeled grey loops in Figure \ref{IntnCCOP}a, where both frequencies are relatively faint. Finally, the short loop represents a loop falling between $L_1$ and $L_0$, for which $\nu_2$ is bright, but $\nu_1$ does not occur in the loop.  

Loops having a given range of lengths can occur at multiple locations within an active region, and they need not have similar orientations. This is indicated schematically in Figure \ref{orientation}a to emphasize that the loops need not be nested as in a single, simple arcade.
The intensity as a function of  distance along the loop for individual loops is shown in Figure \ref{orientation}b \& c for two extremes of relative beam length. Figure \ref{orientation}b shows flattening of the intensity near the loop top for the case where the chosen beam length is shorter than the volume elements, $\Delta s_i > L_b$ as mentioned in section \ref{logistics}, so the intensity near the loop apex is proportional to $L_b$. Figure \ref{orientation}c shows the same plot for a beam length that is longer than the volume elements everywhere along the loop i.e. $\Delta s_i < L_b$ with the intensities proportional to $\Delta s_i$ even near the loop apex.

The corona has a continuous distribution of loop lengths, but we imagine that only a fraction of loops experience nanoflares. We consider three different population densities of such loops. The loop length spacing is $\Delta L / \eta$, where $\Delta L$ is the spacing in Figure \ref{IntnCCOP}a.

\textit{Model 2.1 (Low Loop Population Density)}:  We first examine the case $\eta = 1$, corresponding to a low population density (Figure~\ref{IntnCCOP}a). There are 23 loops in the chosen range of loop lengths. Assuming $500$~nanoflares/s in an entire active region, we estimate $30$~nanoflares/s over this range. All nanoflares are assumed to produce electron beams lasting 20~ms. This is shorter than the typical travel time between the two frequencies.

The nanoflares occur at random times and random locations within the loops. Electron beams are assumed to propagate in both directions. We run the simulation for 1200~s, during which time 36,000 nanoflares are initiated. Figure \ref{IntnCCOP}c shows the CCOPS for the frequency pair $\nu_1-\nu_2$, where we have used all except the first and last 10~s of the light curves. The dashed vertical lines indicate the expected time-lags for loops $L_1,~L_2,$ and $L_3$, color-coded to match the spectra in Figure \ref{IntnCCOP}a. There are clear peaks at the expected locations, especially for $L_1$ (red). It is not surprising that this loop dominates, as it produces the brightest emission at both frequencies, but especially $\nu_1$, which comes from near the apex. The CCOPS peaks are much weaker for $L_2$ (green) and $L_3$ (blue), with $L_2$ being somewhat stronger because it is somewhat brighter.

\begin{figure}[h!]\centering
\includegraphics[ width=1.0 \textwidth, clip]{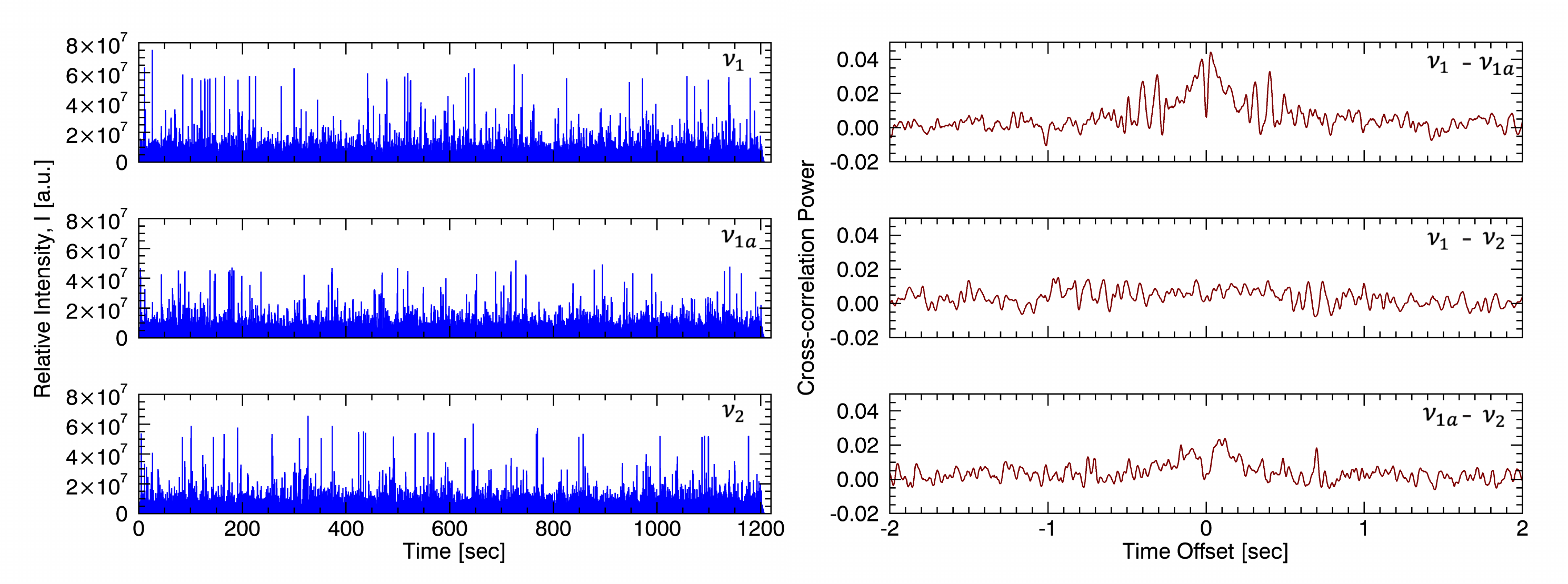}
\caption{\label{LCnCCOPS1151HNF}Left: Light curves for the three chosen frequencies, $\nu_1,~\nu_{1a},~\nu_2$ for high-density loop distribution with 30 nanoflares occurring per second all generating type III bursts. Right: Corresponding CCOPS for each pair.}
\end{figure}

All loops except $L_0$ will produce 4 peaks in the CCOPS, since at least some nanoflares occur below the location of $\nu_2$ in each case. Notice that the short delay, $\pm \Delta t_1$, is smaller for $L_2$-$L_n$ than for $L_1$. Similarly, the long delay, $\pm \Delta t_2$, is bigger. This is easily understood based on the sketches in Figure \ref{difflenloops}, where the intermediate loop represents $L_1$ and the long loop represents $L_2$ and $L_3$. The conjugate frequency positions in the right legs are not shown.

There is a low level of ``noise" in the CCOPS. This is not due to true noise in the light curves, which will be discussed later. Rather, it is due to $\nu_1$ from one nanoflare in one loop correlating with $\nu_2$ from a different nanoflare in a different loop. There is no temporal relationship among the nanoflares, and so the power associated with these ``false" correlations is spread roughly evenly over the range of offsets. Note that the power can be both positive and negative, as expected based on the definition of the cross correlation (see Appendix~\ref{CCOP}).   

\textit{Model 2.2 (Moderate Loop Population Density)}: 
Now consider the situation where the loop population density is three times greater: $\eta = 3$. This is shown in Figure \ref{IntnCCOP}b and d. There are 69 loops in total, spanning the same range of lengths as before. New peaks appear in the CCOPS that were not present previously. We expect three times as many, though most are not visible. The amplitudes of the peaks that are common to both simulations are reduced (lower cross-correlation power), as discussed in Appendix~\ref{CCOP}. 

Notice that the spectra for the two loops between $L_0$ and $L_1$ in Figure \ref{IntnCCOP}b do not have meaningful peaks in the CCOPS because they only emit at frequency $\nu_2$. That emission is quite strong, however, and exacerbates the problem with false peaks. This is a primary reason why the amplitudes of the true peaks are reduced compared to the first simulation.

\textit{Model 2.3 (High Loop Population Density)}: Our final example has a much higher population density that approximates a truly continuous distribution of loop lengths: $\eta = 50$, resulting in 1150 loops. This is essentially equivalent to every field line in a continuous distribution having the opportunity of experiencing a nanoflare. We include a third frequency for the analysis, $\nu_{1a}$, that lies between $\nu_1$~\&~$\nu_2$, marked by the dotted line in Figure \ref{IntnCCOP}a,b. For this simulation we expect a forest of peaks in the CCOPS. Light curves for the three frequencies are shown in Figure \ref{LCnCCOPS1151HNF}, left panel. The associated CCOPS for each pair are shown in the right panel. We note that the power of the cross-correlation has reduced drastically. Even so, a signature is visible above the $3\sigma$ level for at least the pair $\nu_1$--$\nu_{1a}$. There is a noticeable dip at `zero' lag that increases towards both positive and negative time offsets, peaks at a certain lag, and then decreases again. This `M' shaped pattern is the result of a nest of peaks from multiple loops with very similar time-lags. 
The peaks for longer delays corresponding to $\pm~\Delta t_2$ from Figure \ref{sample_CCOP} are visible for a few of the brightest loops only.

A similar pattern is also visible in the CCOPS for the other two pairs of frequencies, but it is important to note here that the signal-to-noise ratio (SNR) of the power is quite low. 

\begin{figure}[h!]\centering
\includegraphics[ width=1.0 \textwidth, clip]{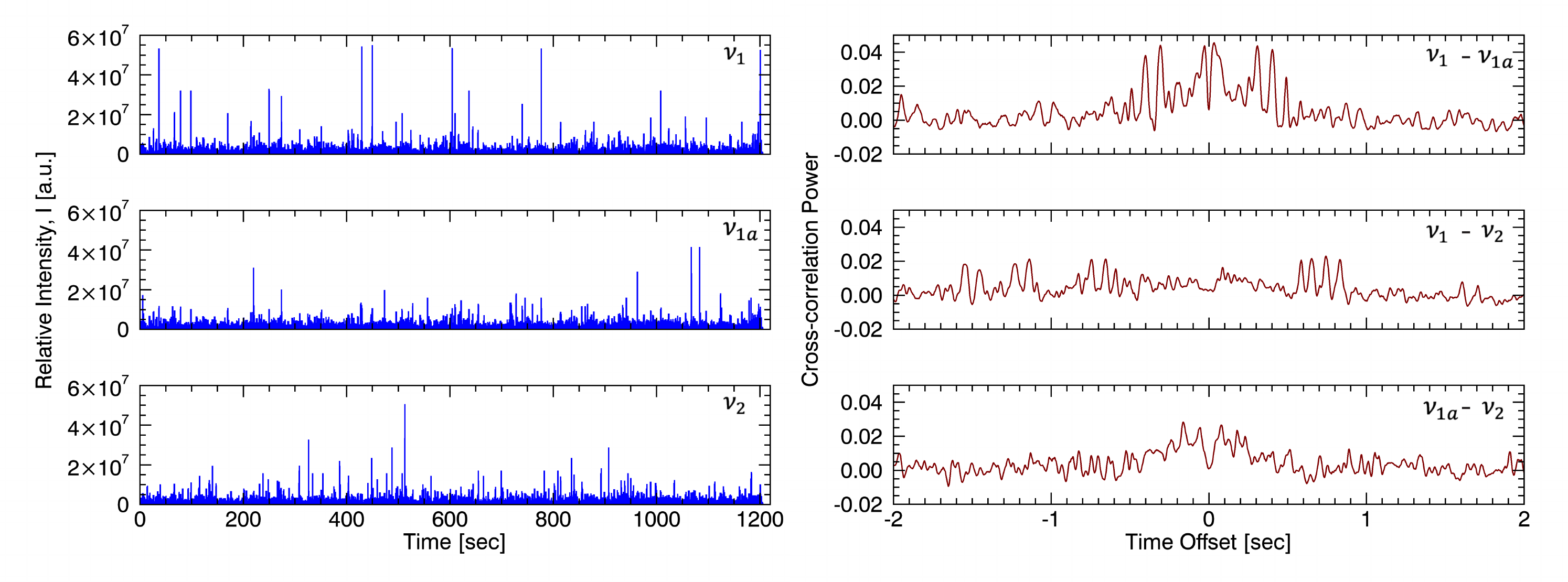}
\caption{\label{LCnCCOPS1151LNF}Left: Light curves for the three chosen frequencies, $\nu_1,~\nu_{1a},~\nu_2$ for the high loop density distribution with 30 nanoflares occurring per second but only a tenth of them generating type III bursts ($3$~bursts/s). Right: Corresponding CCOPS for each pair.}
\end{figure}

The distinctive dip in power at zero lag can be understood on the basis of Figure \ref{difflenloops}. Small lags are produced in loops where both frequencies are close to the footpoint. The intensity is greatly reduced at these locations, so the cross correlation power is weak.

\begin{figure}[h!]\centering
\includegraphics[ width=1.0 \textwidth, clip]{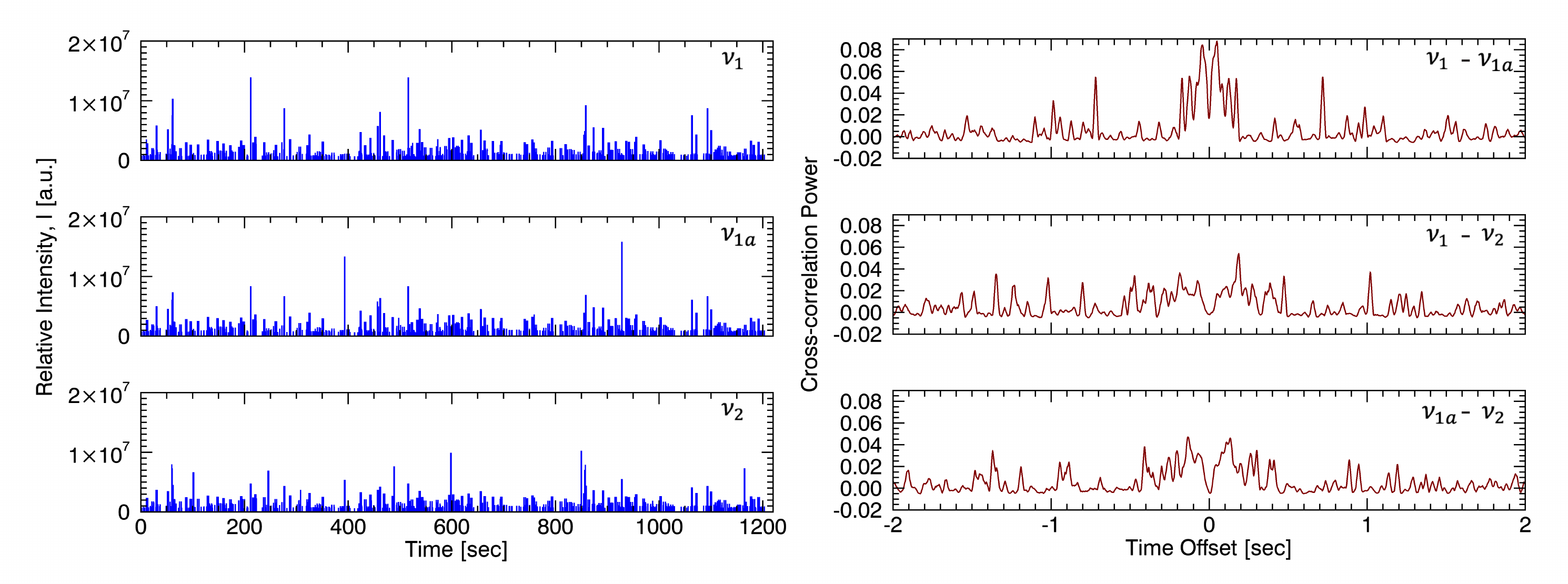}
\caption{\label{LCnCCOPS1151VLNF}Left: Light curves for the three chosen frequencies, $\nu_1,~\nu_{1a},~\nu_2$ for high-density loop distribution with 30 nanoflares occurring per second but only a hundredth of them generating type III bursts ($0.3~bursts/sec$). Right: Corresponding CCOPS for each pair.}
\end{figure}

\begin{figure}[h!]\centering
\includegraphics[angle=0, width=.8 \textwidth, clip]{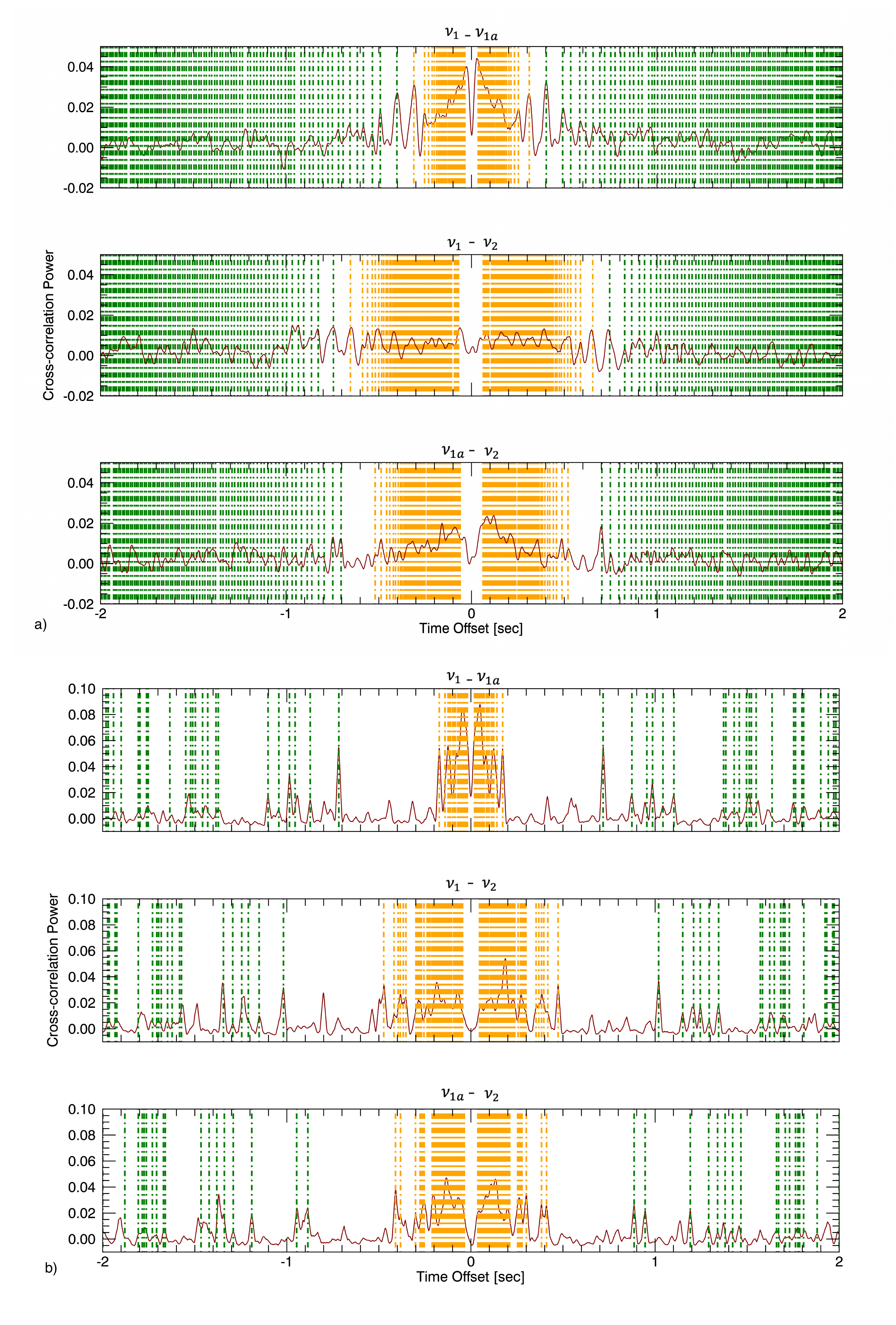}
\caption{\label{peakcomparison}a. CCOPS for 30 bursts/s with expected time-lags overlaid.  b. CCOPS for 0.3 bursts/s with expected time-lags overlaid. The orange dashed lines mark the expect time-lags $\Delta t_1$ and the green dashed lines mark the expected time-lags $\Delta t_2$}
\end{figure}

\subsubsection{Burst-Frequency} \label{burst frequency}
The three simulations above assume that nanoflares occur at a rate of 30 per second across the range of loop lengths considered, and that every nanoflare produces a type III burst. It is certainly possible that only a fraction of nanoflares accelerate energetic electrons, and so the rate of type III bursts could be much less. We therefore repeat $Model~2.3$ above (high loop population density), but with a ten times smaller type III burst rate: $3~bursts/sec$. The light curves for all three frequencies are shown on the left in Figure~\ref{LCnCCOPS1151LNF}, and the CCOPS for three frequency pairs are shown on the right. Note that the CCOPS here are not very different from the ones shown in Figure \ref{LCnCCOPS1151HNF}. An `M' shaped pattern and dip at zero lag are still present. 

The results from a still-further decrease in the burst frequency to one in every hundred nanoflares ($0.3~bursts/sec$) is shown in figure \ref{LCnCCOPS1151VLNF}. 

Figure \ref{peakcomparison} a) shows CCOPS from Figure \ref{LCnCCOPS1151VLNF} and b) shows CCOPS from Figure \ref{LCnCCOPS1151HNF} corresponding to $30~bursts/sec$ and $0.3~bursts/sec$ respectively with the expected time-lags overlaid. The orange dot-dashed lines clearly show how the M-shaped pattern arises from a clustering of expected peaks at the shorter time-lags $\Delta t_1$. The longer time-lags 
$\Delta t_2$ are marked by the green dot-dashed lines. For a high burst frequency (panel a), the longer lags have waned due to the sheer number of bursts, however many of the individual peaks are prominent at the expected positions for the low burst frequency (panel b).

An observation to make is that as we reduce the burst frequency to one-tenth and then further to one-hundredth of the chosen nanoflare rate, there is an increase in power of the cross-correlation peaks. Note that we expect fewer `false' correlations within the range of offsets shown (see Appendix \ref{CCOP}). The average interval between bursts is 3 seconds in the lowest frequency simulation, which is outside the range of -2 to +2 s in the figure. The overall level of ``noise'' is therefore reduced. Nonetheless, some of the distinctive narrow peaks in the CCOPS are false peaks associated with pairs of closely spaced events that occur near the apex of two different loops and are therefore relatively bright at both frequencies.

\begin{figure}[h!]\centering
\includegraphics[ width=1.0 \textwidth, clip]{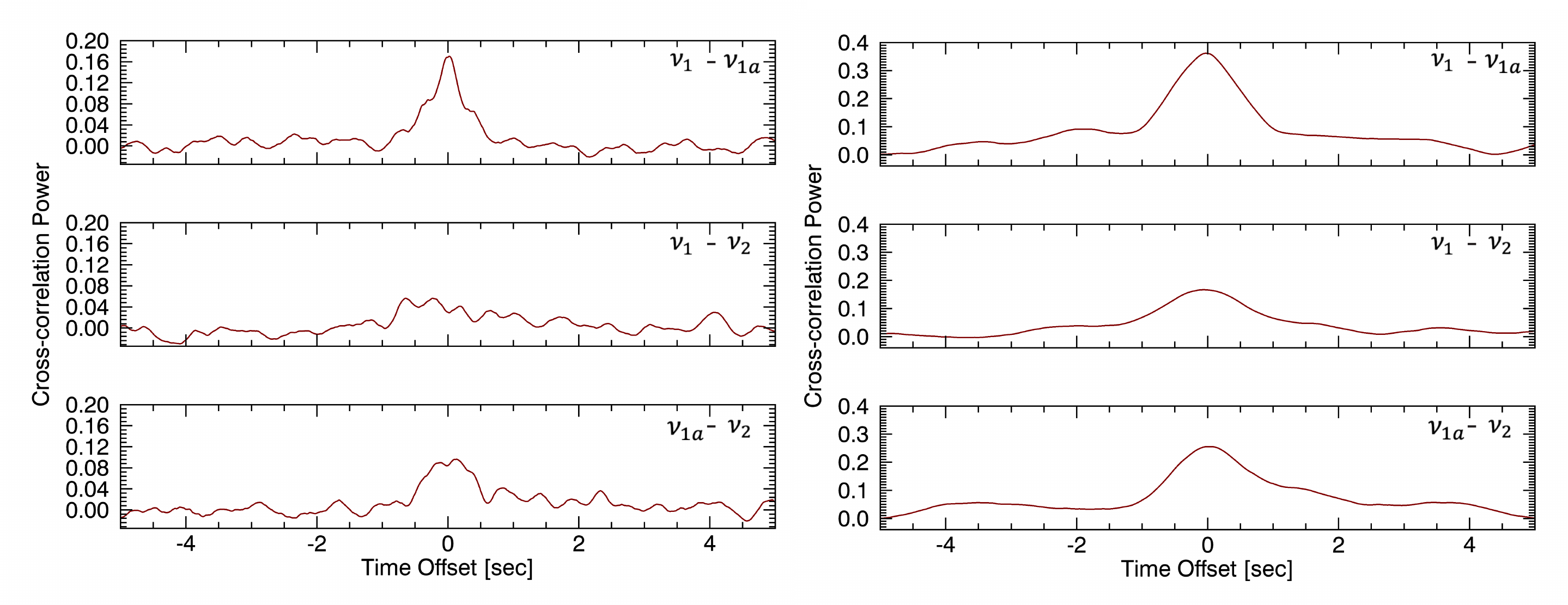}
\caption{\label{duration}Left: CCOPS for the three chosen pairs of frequencies for a burst duration of $200~ms$. Right: CCOPS for the same frequency-pairs for a burst duration of $1~s$.}
\end{figure}

\subsubsection{Duration of the Bursts} \label{burst duration}
Traditional type III bursts in the inner heliosphere may last for a few tens of minutes \citep{2014RAA....14..773R}, however for type IIIs occurring in the closed corona, the observed durations are much shorter \citep{Chen2018}. The burst duration depends on both the lifetime of the acceleration process and any other effects such as the decay time of the Langmuir waves or scattering in the volume. We present our model as though the beam is an emitting object, but it is understood that the duration of emission from a given point in space includes these other effects, i.e., our 'beam' is a column of emission that is longer than the electron beam. The duration of the burst at each frequency $\nu_i$ then depends on the time it takes for the beam to propagate through the $\Delta s_i$ volume element associated with the central frequency $\nu_i$. For all multi-loop models discussed above, the duration of the beam was chosen to be $20$~ms.  Most of the expected time lags between the frequency pairs are longer than this duration.

We now evaluate how well the technique performs when the beam duration is comparable to or longer than the expected time lags. Figure \ref{duration} shows the CCOPS for $Model~2.3$ (comparable to Figure~\ref{LCnCCOPS1151HNF}) except with beam durations of $200~$ms and $1~$s. As the duration increases, the peaks in the CCOPS broaden and begin to merge. The dip at zero lag fills in, and the `M'-shaped pattern disappears, though a hint remains for frequency pair $\nu_{1a}-\nu_2$ at a duration of $200~$ms. The individual peaks and `M'-shaped pattern are broadened to produce a pronounced hump centered at zero lag. Although this removes the direct delay signature of the type III bursts, because such a hump could instead be a result of in-phase variability at the two frequencies, the width and the shape of the hump may still provide evidence for drifting (type III) bursts if its width is substantially greater than the combined durations of the variations in the light curves, as discussed in Appendix B.

\begin{figure}[h!]\centering
\includegraphics[ width=1.0 \textwidth, clip]{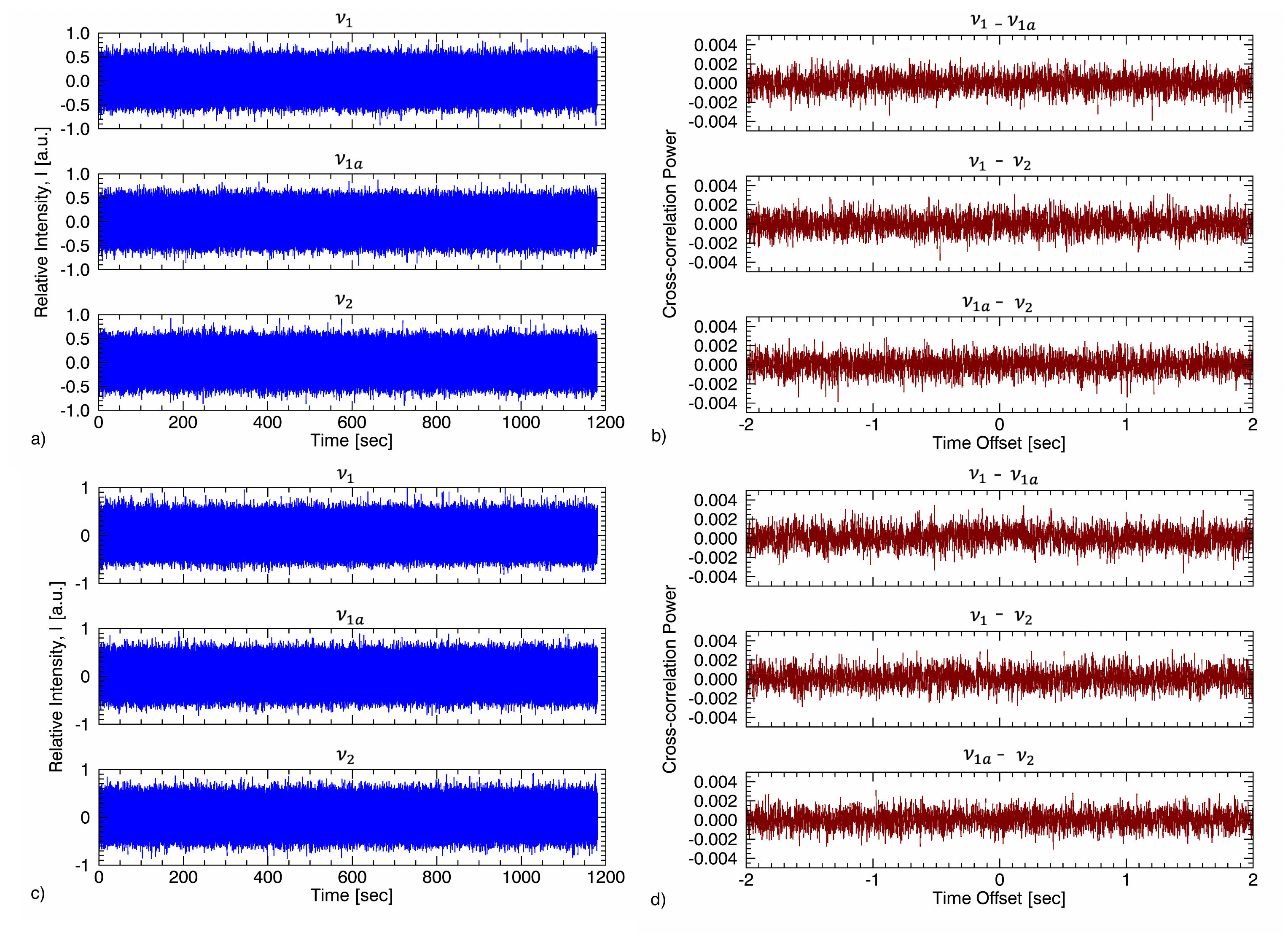}
\caption{\label{noisy1} (a) Flux density for a brightness temperature $T = 10^{12}$ K at frequencies  $\nu_1,~\nu_{1a},~\nu_2$ for the high loop density case with 0.3 bursts/s. Random Gaussian noise with an rms of 0.09 sfu is added to each of the light curves.  (b) The CCOPS corresponding to the light curves in (a).  (c) Same as (a) for a higher rate of 30 bursts/s.  (d) The CCOPS corresponding to the light curves in (c).}
\end{figure}

\subsubsection{The Role of Noise} \label{noise}

The final test is to model the technique with an additional level of noise. For this purpose, we use $Model~2.3$ again. To add a realistic level of noise to the light curves we need to understand the various factors that affect the sensitivity of a radio interferometer. At frequency $\nu$, the flux density associated with a given antenna temperature\footnote{This is not the antenna's physical temperature but the temperature that a blackbody would have in order to provide the equivalent power received by the antenna.} is given as:

\begin{equation}\label{eq6}
   S = \frac{2kT_a\nu^2}{c^2}d\Omega,
\end{equation}

\noindent where $k$ is the Boltzmann constant, $T_a$ is the antenna temperature due to a hot source filling the antenna primary beam (field of view), $c$ is the speed of light, and $d\Omega$ is the solid angle subtended by the beam. If a hot source of temperature $T$ does not fill the beam, the antenna temperature is reduced ($T_a < T$) by the beam dilution factor (ratio of the angular area of the source to the beam area). The noise associated with a radio antenna is a combination of this antenna temperature due to the source and the noise generated internally by the receiving system, $T_{\rm sys}$. \cite{Crane1989} showed that the sensitivity for an array of such antennas is given by:

\begin{equation}\label{eq7}
   \Delta T = \left[\frac{T_a^2 + T_aT_{\rm sys} + {T_{\rm sys}}^2/2 }{\Delta \nu \tau N(N-1)}\right]^{1/2},
\end{equation}

\noindent where $\Delta \nu$ is the channel width of the instrument and $\tau$ is the cadence, and $N(N-1)$ is the number of baselines. For a bright source such as the Sun, $T_a >> T_{\rm sys}$. Simplifying Eq. (7) and using Eq. (6), we have the noise associated with a source of flux density $S$ as

\begin{equation}\label{eq8}
   \Delta S = \frac{S}{\sqrt{\Delta \nu \tau N(N-1)}}.
\end{equation}

Therefore, for a source with flux density $S$, the sensitivity has a direct dependence on $T_a$ and an inverse dependence on the frequency bandwidth $\Delta \nu$, integration time $\tau$, and number of baselines.

\begin{figure}[h!]\centering
\includegraphics[ width=1.0 \textwidth, clip]{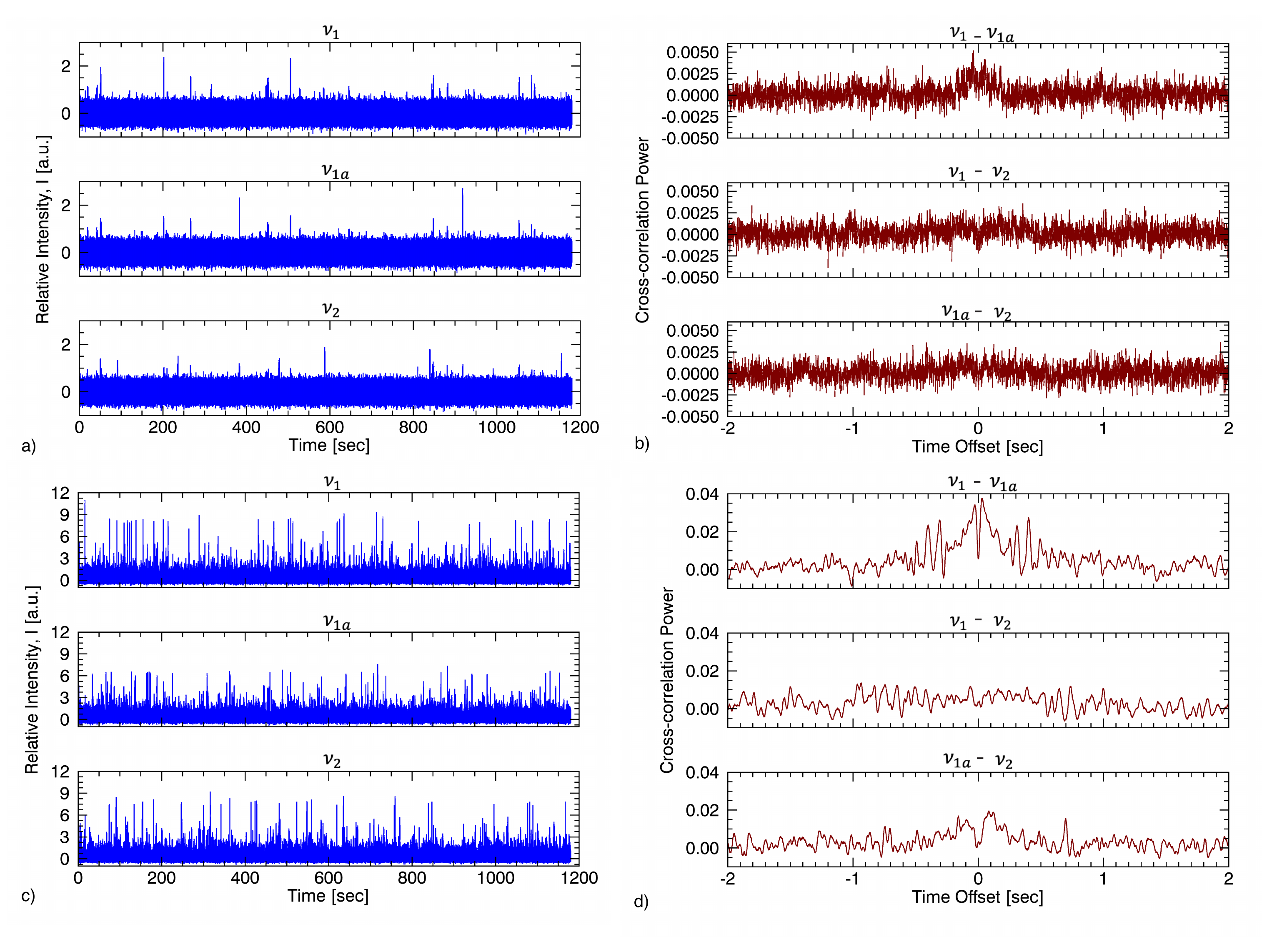}
\caption{\label{noisy2} Same as Figure~\ref{noisy1} for bursts with an order of magnitude higher brightness temperature, $T=10^{13.3}$ K.}
\end{figure}

\begin{figure}[h!]\centering
\includegraphics[ width=1.0 \textwidth, clip]{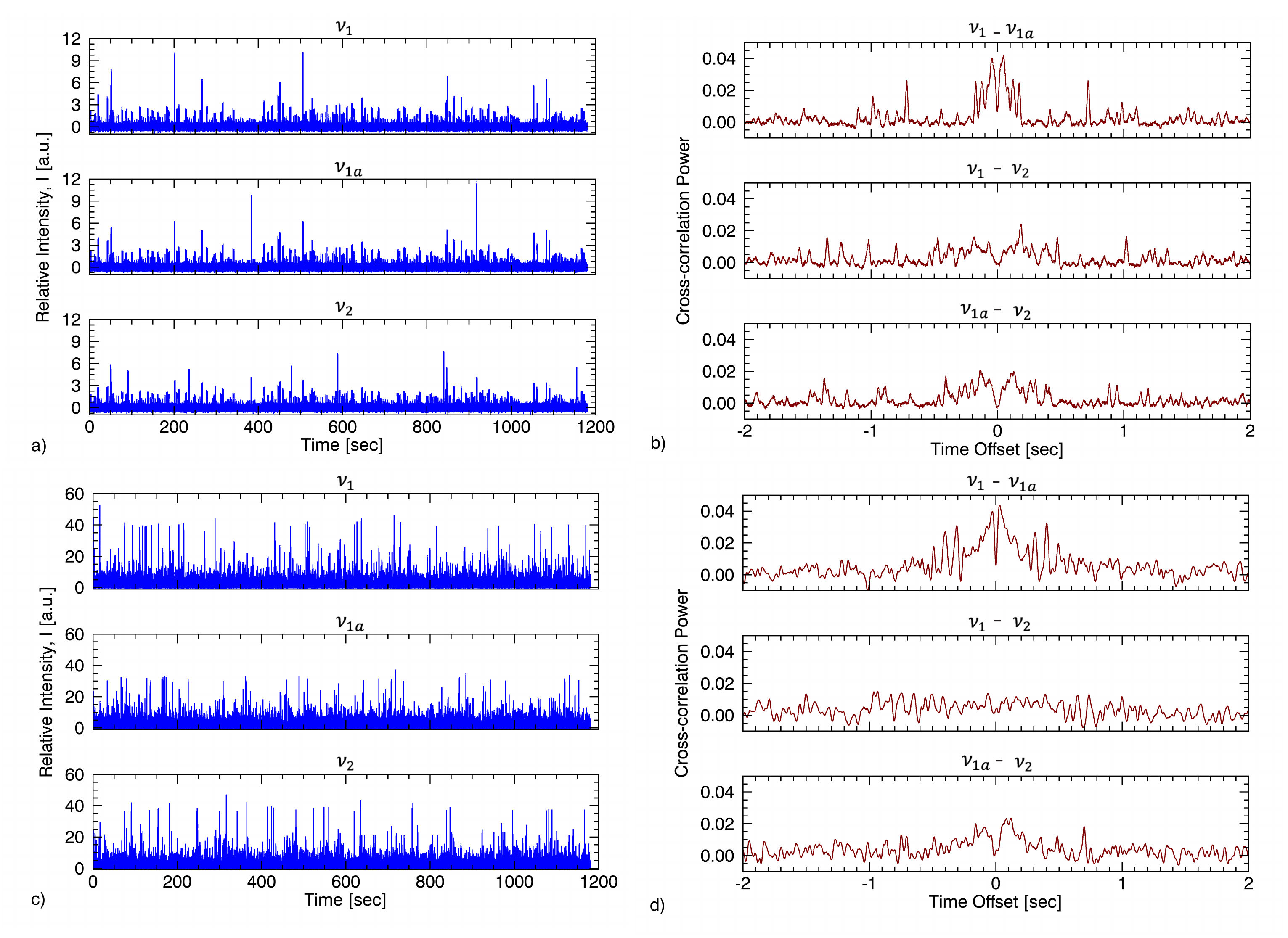}
\caption{\label{noisy3} Same as Figure~\ref{noisy1} for bursts with three orders of magnitude higher brightness temperature, $T=10^{14}$ K.}
\end{figure}

 Anticipating a future analysis of $P$-band (245--450 MHz) solar observations we have in hand, taken with the Very Large Array (VLA)\citep{Perley2011}, we estimate a level of noise as follows. The calibration procedure for $P$-band data from the period of interest has not yet been finalized, so we adopt an average flux of 25 sfu at 245 MHz based on values reported in the NOAA catalogue for times of similar activity. Using numbers appropriate to our existing $P$-band observations, $\Delta \nu = 125$ kHz, $\tau = 0.01$ s and number of antennas in the subarray $N = 14$, equation \ref{eq8} gives an uncertainty of $\Delta S = 0.0524$ sfu. 
 
 An independent estimate of the uncertainty can be obtained from the data by taking the root mean square (RMS) intensity difference between two frequencies that are so closely spaced that any real differences in source flux density are likely to be negligible, although fluctuations due to source confusion \citep{1978ApJ...224.1043Z} may still be present. Because of possible cross-talk between adjacent channels, we use channels separated by one intervening channel. Lacking calibrated $P$-band data, we use data near 1 GHz from the $L$-band, which are calibrated. The RMS difference is 3.5 times larger than the uncertainty given by Equation \ref{eq8} using the values appropriate to that band. This suggests the presence of additional sources of random variations, perhaps akin to those reported by \cite{1978ApJ...224.1043Z}. We assume that they are also present at about the same level in the $P$-band data, so we multiply the 245 MHz uncertainty above by the same factor of 3.5 to obtain $\Delta S =0.186$ sfu, which we apply to our models.

All intensities from $Model~2.3$ are first converted to a flux density using equation \ref{eq6} above, where the solid angle $d\Omega$ is calculated for each nanoflare at frequency $\nu$ with a volume element $\Delta s$ and assuming a width of $\approx 200$ km for the magnetic strand \citep{2015RSPTA.37340256K}. Since the brightness temperatures of these nanoflares are unknown, we repeat the calculation for three temperatures, viz. $10^{12}$, $10^{13.3}$ and, $10^{14}$~K. Randomly generated Gaussian noise with an rms of 0.186 sfu is then added to the light curves of each frequency and the CCOPS computed.

Figures \ref{noisy1}, \ref{noisy2} and \ref{noisy3} show flux densities computed at temperatures $T = 10^{12}, 10^{13.3}$ and $10^{14}$ K at all three frequencies for 0.3~bursts/s and 30~bursts/s in panels a and c, respectively, with the noise added. The corresponding CCOPS for each pair of frequencies are shown in panels b and d. Light curves in Figure \ref{noisy1} computed for temperature $T = 10^{12}$ K are completely dominated by noise and show no meaningful peaks in the CCOPS. 
The rather specific value $T = 10^{13.3}$ K was chosen as an example where there are only minimal excursions above the noise in the light curves  (Fig. \ref{noisy2} a), yet the CCOPS (panel b) shows a hint of the `M'-shaped pattern. The flux densities of the higher burst frequency case in panel c have a substantial increase in the signal above the noise, and hence the CCOPS now look very similar to Figure \ref{LCnCCOPS1151HNF} with a very slight decrease in the power.

Any further increase in the brightness temperature significantly increases the signal-to-noise ratio for both cases of burst frequencies and consequently resuscitates the CCOPS to their former shapes with a slight reduction in the cross-correlation amplitude that is caused by the noise. The light curves and corresponding CCOPS for such a case with $T = 10^{14}$ K are shown in Figure \ref{noisy3}.Despite such an increase in the brightness of the bursts, the fraction of total flux density due to nanoflares is only 0.055.

\section{Conclusion} \label{discussion}
We perform numerical modeling to simulate idealized emission from type III radio bursts that may be generated by particle acceleration from nanoflares. For the sake of simplification, our model makes the following assumptions: (i) our model loops are symmetric; (ii) all bursts produce electrons of the same velocity, which remains constant as the beams propagate along the loops; (iii) the burst emission is generated instantaneously as the beam is ejected from the nanoflare site; (iv) the emissivity for all bursts is taken to be the same at all loop positions; (v) the decay time of the Langmuir waves is independent of frequency. 

Relaxing assumption (i), loop symmetry, would affect the time delay symmetry shown in Figure~\ref{sample_CCOP}. Regarding assumption (ii), modeling of type IIIs by \cite{Reid2018} shows that the velocity of the bursts eventually decreases as the beam moves away from the Sun, however the evolution of beam speeds in closed loops is unknown. Assumption (iii) ignores the fact that it may take time for the beam to develop a two-stream instability and hence produce Langmuir waves, but this should not affect our results unless nanoflares occur strongly preferentially at the apex of loops, since we have shown that emission on closed loops is expected to be dominated by loop-top emission from beams originating at any location. Assumption (iv) would alter the intensity curves of Figure~\ref{int_cartoon}, but the intensity dependence is so strong that only extreme violations of this assumption in favor of loop leg emission would make much difference. Assumption (v) should not be significant unless the decay time has a strong dependence on frequency, since only closely spaced frequencies contribute significantly to the CCOPS. Some aspects of assumption (v) were explored in our study by showing how different durations of bursts will affect the CCOPS.

Once the light curves are obtained from our model, the simulated light curves at chosen radio frequencies are then cross-correlated to identify time-lags between different pairs, using a novel application of the \cite{2012ApJ...753...35V} time-lag technique. We find that the signature of the bursts depends very much on the rate and duration of nanoflares and on the fraction of loops that are involved. Individual peaks dominate the CCOPS when only a small subset of loops experience nanoflares that accelerate energetic particles. When many loops experience such nanoflares, the signature varies depending on whether the beam duration is short or long compared to the particle travel time associated with the two frequencies. Short durations produce a quasi-continuous `M'-shaped pattern with a distinctive dip at zero lag. Long durations produce a broad hump centered at zero lag. These differences can be exploited to determine the likelihood of particle acceleration and the properties of the beam, and therefore better understand the acceleration mechanism.

In general, the signatures are stronger for pairs of frequencies that are closely spaced, indicating that high frequency resolution observations must be used. This is due to the extremely steep slope of the type III spectrum of an individual loop, which is related to the highly nonuniform density gradient along the loop. In order for the intensities to be reasonably strong, both frequencies must occur relatively high in the loop. Emission from the lower leg and transition region is comparatively negligible. 

False peaks can appear in the CCOPS, especially if the burst rate is high. These are due to correlations between different nanoflares occurring in different loops. The value of the time lag of such peaks is not meaningful, but their mere existence would be indicative of radio bursts and therefore would be an important observational feature. 
Whether type III bursts from nanoflares are detectable depends on the SNR. Our noise tests based on estimates of noise for VLA data reveal that for type IIIs with a low brightness temperature, the noise will completely wash out any emission and the CCOPS will show no correlation. Given a high enough brightness temperature for the events, the emission does rise above the noise level and although the power of the cross-correlation peaks in the CCOPS remains small, the change in the widths of the peaks in comparison to the CCOPS computed for a noise-dominated simulation itself is an indication of the presence of bursts. We note that for an instrument with higher sensitivity compared to VLA, the power of the peaks in the CCOPS will improve.

A possible criticism of the central idea of this paper, that nanoflares may produce type III emission on closed loops, lies in the apparent fact that observations provide little evidence for type III emission on closed loops beyond relatively rare `U' or `J' bursts.  However, our analysis of the expected emission from equilibrium loops reveals that such emission should be extremely strongly concentrated at the loop apex (Figure~\ref{int_cartoon}c). Remarkably, this predicts that the observational signature of type III bursts in closed loops is a bright, narrow-band feature with perhaps a faint high-frequency tail, which fits the description of the very commonly observed type I bursts. The narrow spread is expected if the burst occurs in a single loop or several adjacent loops with similar apex densities, and the emission frequency would directly provide the loop apex density. Thus, our model suggests a possible origin of type I bursts as a natural consequence of type III emission in a closed-loop geometry.  We emphasize that the extreme variation in brightness is a consequence of the density profile that we expect in a closed-loop geometry. This modeling approach, however, is applicable to other density profiles, such as those in flux tubes that are part of magnetically open coronal holes and the solar wind. As we will report in a future paper, the type III brightness variation is much more moderate in this case, both in frequency and spatial location. We suggest that classic `U'  and `J' bursts occur on open flux tubes with significant bend or on giant closed loops where the constant conductive flux model is no longer appropriate. Finally, we note that \cite{2011A&A...526A.137D} have suggested that both distinguishable type III bursts and noise storms can be produced together by interchange reconnection at the boundary between open and closed field lines. To our knowledge, we have offered the first explanation for why the type Is have a narrow frequency spread. 

We have begun to apply the knowledge gained from these simulations to actual observations. Results from ground-based radio observatories and the FIELDS instrument on Parker Solar Probe will be reported in upcoming papers.

\acknowledgments

We thank Nicholeen Viall for valuable discussions on various aspects of the time-lag technique and Golla Thejappa for his constructive comments and valuable insights in understanding the theoretical aspects of type III bursts. We also thank the reviewer for their thorough review and comments. This work was supported in part by the NSF grants AST-1615807, AGS-1654382, and AST-19010354; the NASA grants 80NSSC18K1128 and 80NSSC17K0660 to New Jersey Institute of Technology; the Internal Scientist Funding Model (competitive work package program) at Goddard Space Flight Center; and the FIELDS experiment on Parker Solar Probe.

\appendix \label{appendix}

\section{Type III intensity variation} \label{intensity}
Under the assumption that the intensity, $I$ for a constant frequency bin $\Delta \nu$ along the loop is directly proportional to the volume $\Delta s$ that the particles traverse, we have: 
\begin{equation}\label{a1}
    I  \propto \Delta s = \frac{\Delta n}{dn/ds}
\end{equation}
where $\Delta n$ is the density bin corresponding to the frequency bin $\Delta \nu$.

The plasma frequency, $\nu \propto n^{1/2}$, which gives us:
\begin{eqnarray}\label{a2}
      \Delta \nu & = & \frac{d\nu}{dn}\Delta n %\\
    % & \propto & \frac{1}{2}n^{-1/2}\Delta n  \nonumber\\
    % & \propto & \frac{1}{2}\nu^{-1}\Delta n \nonumber
\end{eqnarray}

\begin{equation}\label{a3}
   \Rightarrow \Delta n \propto 2\nu \Delta \nu   
\end{equation}
\begin{equation}\label{a4}
    I \propto 2\nu \left(\frac{dn}{ds}\right)^{-1}\Delta \nu
\end{equation}
Note that, as mentioned in Section \ref{logistics}, this relation is true for all frequencies $\nu_i$ for which the respective volume element is smaller than the beam length, i.e. $\Delta s_i < L_b$. For any element $\Delta s_i > L_b$, the intensity, $I$ is directly proportional to the beam length, $L_b$.
The intensity thus calculated is used to create the light curves for each frequency based on their position along the loop.

To visualize the above relation better, let us look at it in terms of temperature, $T$ along the loop. From equation \ref{a4}, for a constant volume element $\Delta \nu$, we have,
\begin{eqnarray}\label{a5}
    I & \propto & \nu \left(\frac{dn}{ds}\right)^{-1}\\
    & \propto & n^{1/2} \left(\frac{dn}{ds}\right)^{-1}  \nonumber
\end{eqnarray}

 The structure of a static equilibrium loop is reasonably well represented by a constant conductive flux at constant pressure. For a constant conduction flux $F_c$

\begin{eqnarray}
    F_c & \propto & T^{5/2} \left(\frac{dT}{ds}\right)\\
    & \propto &T^{5/2} \frac{d}{ds}\left(\frac{P}{n}\right)  \nonumber\\
    & \propto &T^{5/2} \left(-\frac{1}{n^2}\right)\frac{dn}{ds}  \nonumber
\end{eqnarray}
assuming constant pressure along the loop.

\begin{equation}
   \Rightarrow \frac{dn}{ds} \propto T^{-5/2}n^2 
\end{equation}
 
  Substituting this back in equation \ref{a5}, we get
 
 \begin{eqnarray}
    I & \propto & T^{5/2}~n^{-2}~n^{1/2}   \\
    & \propto & T^{5/2}~T^{2}~T^{-1/2} \nonumber
\end{eqnarray}\label{a9}
\begin{equation}
   \Rightarrow I \propto T^{4} \propto n^{-4} \propto \nu^{-8}
\end{equation}
The above equations clearly exhibit the precipitous fall in intensity $I$ as the density $n$ increases downward along the loop and the $T$ decreases.

\section{Factors affecting the CCOP} \label{CCOP}
The basis of the time-lag technique \citep{2012ApJ...753...35V, 2017ApJ...842..108V} is to compute the cross-correlation between a pair of light curves for two different channels as a function of imposed temporal offset, $l$. For some studies, such as the cooling of nanoflare-heated loops, only the lag of maximum correlation is important. For the purpose of this study, we are interested in the complete spectrum of cross-correlation power. The equation used to calculate the cross-correlation power, $P$ for a negative lag is given as:

\begin{equation}\label{b10}
    P(l<0) = \frac{\sum_{k=0}^{N-|l|-1}(x_{k+|l|}-\bar x)(y_k - \bar y)}{\sqrt{\sum_{k=0}^{N-1}(x_k-\bar x)^2 \sum_{k=0}^{N-1}(y_k-\bar y)^2}}
\end{equation}

For a positive lag, 

\begin{equation}\label{b11}
    P(l>0) = \frac{\sum_{k=0}^{N-l-1}(x_k-\bar x)(y_{k+l} - \bar y)}{\sqrt{\sum_{k=0}^{N-1}(x_k-\bar x)^2 \sum_{k=0}^{N-1}(y_k-\bar y)^2}}
\end{equation}
where $x_k$ and $y_k$ are the intensities in the chosen frequencies at time $k$, $\bar x$ and $\bar y$ are the means of the intensity light curves for $x$ and $y$ respectively, and $N$ is the number of data points in the light curves.

\begin{figure}[h!]\centering
\includegraphics[ width=1.0 \textwidth, clip]{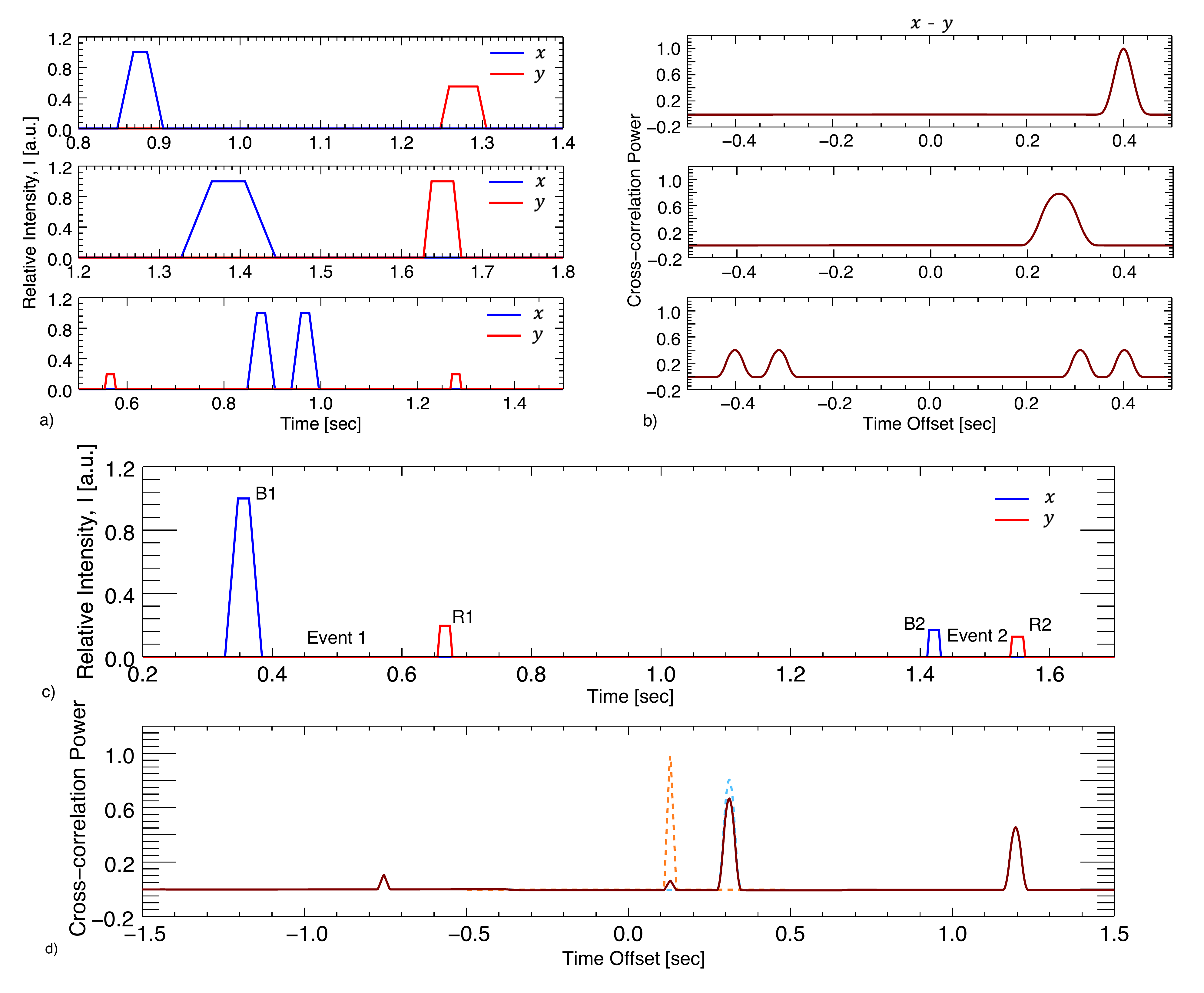}
\caption{\label{appendixfig}a) Light curves for two frequencies $x$ and $y$ overlaid on one plot for three different scenarios; Top: The event has the same duration but differs in intensity for the two frequencies, Middle: the event has same intensities but different durations and, Bottom: Both intensities and durations differ, but now two peaks are present at each frequency as the particles move in both directions (same as Case C in Section \ref{single loop}). b) CCOPS corresponding to each scenario from a). c) Light curves for frequencies $x$ and $y$ overlaid on one plot showing two events, Event 1 and Event 2, on different loops (with electron beams moving downward only). d) CCOPS corresponding to then scenario from c). The overlaid dashed line shows individual CCOPS for Event 1 in the absence of Event 2 and vice versa.}
\end{figure}

Now, following from equations \ref{b10} \& \ref{b11} note that the product of deviations from the mean, i.e. $(x_i-\bar x)(y_i-\bar y)$ will be positive only when $x_i$ and $y_i$ both lie on the same side of their respective means. Therefore, if $x_i$ and $y_i$ are simultaneously greater or less than their respective means, the cross-correlation power $P(l)$ will be positive, and if one is greater while the other is less, the power will be negative. While summing over the product for all data points, the negative products will contribute in reducing the power of the cross-correlation.This effect is quite apparent from the equations above. However, additional factors such as the duration and number of the bursts also affect the power of the cross-correlation. 

Let us consider example light-curves for $x$ and $y$ as shown in Figure \ref{appendixfig}a. Each panel in the plot represents a key difference between the light curves. In the top-most panel, the two light curves vary in intensity but the duration of the bursts is exactly the same. In the middle panel, the duration varies but intensities are the same, and in the bottom panel, along with the variation in intensity and duration, we introduce emission in two places. The last case is very similar to the one seen in Case C from Section \ref{single loop} (particles moving in both directions). To compute the CCOP, imagine moving one of the light curves with respect to the other until their centers are perfectly aligned and then moving away in the opposite direction.
As the peaks in the light-curves start overlapping, the power of the cross-correlation increases until it reaches a maximum when their centers are perfectly aligned and then reduces again.

Figure \ref{appendixfig}b shows the CCOPS corresponding to each panel in \ref{appendixfig}a. For the top panel, the maximum power of the cross-correlation is unity at the time-difference required to perfectly align the two light-curves. Their difference in intensity does not affect the CCOP. To see this, imagine that one of the peaks becomes enhanced. The deviation from the mean is then larger throughout the light curve. It is positive and substantially larger at the location of the peak, and it is negative and minimally larger outside the peak. The net effect is no change in the CCOP.

In the middle panel, the maximum power never reaches one. This is simply due to the different durations of bursts in the two light curves, even when they are centered one above the other. Also note that the width of the cross correlation curve is equal to the sum of the burst durations in the two frequencies. In the bottom panel, the maximum power for each time-lag has become much smaller. As one intensity peak in $x$ is centered above one of the peaks in $y$, the misalignment in the other peaks of the light-curves further reduces the power in addition to the different durations.

Lastly, panel \ref{appendixfig}c shows light curves for frequencies $x$ and $y$ for two events. Each event occurs in a different loop and the particles only move downward. The physically interesting correlations are those between peaks of the same event, i.e. $B1$ with $R1$, and $B2$ with $R2$.  Correlations of one burst with another, i.e. $B1$ with $R2$ and $B2$ with $R1$, also exist but are what we term ``false" peaks.  The corresponding CCOPS (solid line) in panel d also have the dashed line overlaid showing individual CCOPS for Event 1 (cyan) in the absence of Event 2 and vice versa (orange). The CCOPS (solid line) show two important implications of the presence of multiple bursts in the light curves: 
\begin{enumerate}
    \item Contrary to the first example (top panel) in panel a), the high intensity from event 1 in $x$ will shift the mean $\bar x$ such that the
    faint peak $B2$ in $x$ will have a 
    smaller positive deviation from the mean. Consequently, the CCOPS will now be affected by the difference 
    in intensity between the two events. The CCOPS peak at $\sim0.31$~s with the highest power corresponds to a cross-correlation between $B1$ and $R1$. Note that its magnitude is lower than it would have been in the absence of Event 2. Also note that the power of cross-correlation between$B2$ and $R2$ at an offset of $\sim0.22$~s is far smaller than its unity value in the absence of Event 1.
    
    \item The second brightest peak in the CCOP, at a time offset of $\sim1.2$~s, corresponds to a cross-correlation between $B1$ and $R2$, i.e. events from two different loops, and hence is a ``false'' peak. The second false peak at the offset of $\sim -0.75$ s from $R1$ and $B2$ also has a small power due to the reason mentioned in the previous point.

\end{enumerate}
This tells us that the CCOP in the presence of multiple bursts will be affected by the dominant intensity peaks in the light curves as they skew the value of mean; and that the same dominant peaks are also responsible for a high CCOP at false time--lags. This effect is enhanced with an increase in the burst frequency.

\bibliography{main.bib}{}
\bibliographystyle{aasjournal}

\end{document}